\documentclass{aa}

\usepackage[percent]{overpic}
\usepackage[export]{adjustbox}

\defcitealias{2025A&A...695A.220K}{Paper I}
\defcitealias{mapping-disk}{Paper II}
\defcitealias{mapping-bulge}{Paper III}
\defcitealias{Mapping-extragalactic}{Paper IV}
\defcitealias{Mapping-sfh}{Paper V}
\defcitealias{Mapping-migration}{Paper VI}

\usepackage[breaklinks=true,colorlinks,citecolor=blue]{hyperref}
\usepackage{graphicx,animate}
\usepackage{enumerate}
\usepackage{epsfig}
\usepackage{color}
\usepackage{txfonts}
\usepackage{natbib}
\usepackage{multirow}
\usepackage{ulem} 
\usepackage{amssymb}

\usepackage{movie15}

\usepackage{booktabs}
\usepackage{array}   

\usepackage{academicons}

\newcommand{\orcid}[1]{\href{https://orcid.org/#1}{\textcolor[HTML]{A6CE39}{\aiOrcid}}}
\usepackage{orcidlink}

\usepackage{xcolor}

\defcitealias{2025A&A...695A.220K}{Paper I}
\defcitealias{Mapping-disk}{Paper II}
\defcitealias{Mapping-bulge}{Paper III}
\defcitealias{Mapping-extragalactic}{Paper IV}
\defcitealias{Mapping-sfh}{Paper V}

\definecolor{ao}{rgb}{0.0, 0.5, 0.0}

\newcommand{\Gaia}{{\it Gaia}\,}

\newcommand{\FeH}{\ensuremath{\rm [Fe/H]}\,}

\newcommand{\Rb}{\ensuremath{\rm R_{birth}}}

\defcitealias{Mapping-1}{Paper I}
\defcitealias{Mapping-2}{Paper II}
\defcitealias{Mapping-3}{Paper III}
\defcitealias{Mapping-4}{Paper IV}
\defcitealias{Mapping-5}{Paper V}

\definecolor{ourdarkblue}{rgb}{0.0, 0.0, 0.55}

\begin{document}

\title{Rediscovering the Milky Way with an orbit superposition approach and APOGEE data V. The disc growth and history of star formation}

\titlerunning{Milky Way growth and SFH}
\authorrunning{Ratcliffe et al.}

\author{Bridget Ratcliffe\thanks{bratcliffe@aip.de}\inst{1}, Sergey Khoperskov\inst{1,2}, Nathan Lee\inst{3}, Ivan Minchev\inst{1}, Paola Di Matteo\inst{4}, Glenn van de Ven\inst{2}, \\Misha Haywood\inst{4}, Léa Marques\inst{1,5}, John Paul Bernaldez\inst{6,1}, Davor Krajnović\inst{1}, Matthias Steinmetz\inst{1} }

\institute{Leibniz-Institut für Astrophysik Potsdam (AIP),
              An der Sternwarte 16, 14482 Potsdam, Germany
              \and
              Department of Astrophysics, University of Vienna, T\"urkenschanzstrasse 17, A-1180 Vienna, Austria
              \and 
              BIFOLD, Berlin Institute for the Foundations of Learning and Data, Berlin, Germany
              \and 
              LIRA, Observatoire de Paris, Université PSL, Sorbonne Université, Université Paris Cité, CY Cergy Paris Université, CNRS, 92190, Meudon, France
              \and 
              Universität Potsdam, Institut für Physik und Astronomie, Karl-Liebknecht-Str. 24-25, 14476 Potsdam, Germany
              \and 
              Friedrich-Schiller Universit\"at Jena, 07737 Jena, Germany\\
              }

\date{Received \today; accepted ...}

\abstract{The Milky Way’s (MW's) star formation history (SFH) offers insight into the chronology of its assembly and the mechanisms driving its structural development. 
}{In this study, we present an inference and analysis of the spatially resolved SFH and the MW disc growth.}{Our approach leverages both stellar birth radii estimates and the complete reconstruction of the MW stellar disc using a novel orbit superposition method from APOGEE data, allowing us to trace the orbit-mass weighted SFH based on formation sites while taking into account stellar mass loss.}{
We find that the MW is a typical disc galaxy exhibiting inside-out formation: it was compact at $z > 2$ ($\rm R_{\rm eff} \approx 2$ kpc), had a peak in its star formation rate (SFR) 9--10 Gyr ago, and grew to a present-day size of $\rm R_{\rm eff} \approx 4.3$ kpc. A secondary peak in SFR $\sim 4$ Gyr ago is responsible for the onset of the outer disc, which comprises the metal-poor, low-$\alpha$ population. We find that in-situ star formation in the solar neighbourhood started 8--9~Gyr ago. The MW disc is characterised by a negative mean age gradient, as the result of  the inside-out growth, with additional flattening induced by stellar radial migration.}{Our work showcases the importance of accounting for radial migration and stellar sample selection function when inferring the SFH and build-up of the MW disc. 
}
   

\maketitle


\section{Introduction}

Star formation is a fundamental process in the formation and evolution of galaxies, not only converting gas into stars but also driving chemical enrichment and contributing to feedback processes that shape the interstellar medium (ISM). A galaxy's star formation history (SFH) reflects the evolution of the star formation rate (SFR), providing insight into merger history, gas accretion, stellar feedback, and secular evolution at different epochs \citep{1980FCPh....5..287T, 1998ARA&A..36..189K,  2012ARA&A..50..531K}. Thanks to extragalactic surveys, the trend in SFR is available across time for a large number of galaxies, providing constraints on how galaxies acquire their stellar mass in different environments as a function of redshift (e.g., \citealt{2011ApJ...742...96W, 2015A&A...579A...2I, 2016ApJ...832...79P, 2023MNRAS.522.6236T, 2023MNRAS.520.3974C, 2025A&A...696A.159A}). Specifically, extragalactic studies revealed a so-called downsizing process reflecting an inversion between the mass hierarchy of galaxy formation and that of stellar assembly. The dark matter builds up hierarchically, and baryonic processes of star formation and chemical enrichment appear to occur earlier and more rapidly in more massive galaxies, with a peak in star formation at redshift z $\sim 2$ \citep{1998ApJ...498..106M, 2005ApJ...619L.135J, 2014ARA&A..52..415M}.

While the causality is not obvious, the evolution of the SFR seems to reflect the morphological diversity of massive disc galaxies~\citep[e.g.,][]{2017A&A...597A..97M}. In particular, among galaxies with similar stellar mass at z $\sim 0$, those that assembled their stellar mass more rapidly are more likely to develop bars, whereas those with more extended or delayed SFHs tend to remain non-barred~\citep{2020MNRAS.499.1116F, 2024MNRAS.533.3975K}. This correlation either suggests that the rapid disc settling favours the bar instability at early times, with the subsequent suppression of star formation by bars, or suggests slowly growing galaxies remain relatively gas-rich with low-mass stellar discs preventing bar formation. Additionally, spatially-resolved studies of integral field unit data suggest that most massive disc galaxies exhibit inside-out growth: central regions formed earlier and outer regions formed over longer timescales ~\citep{2013ApJ...764L...1P, 2014A&A...562A..47G, 2016A&A...594A..36S}. Consequently, this results in negative age and metallicity gradients in most spiral galaxies, weakly depending on merger history and feedback~\citep{2017MNRAS.466.4731G, 2017MNRAS.469..151B, 2017MNRAS.472.2833S, 2018MNRAS.480.2544R, 2018MNRAS.476.3883L}.



The Milky Way~(MW) offers a unique opportunity to study its SFH, as it provides access to individual stars with well-constrained ages, kinematics, and chemical abundances. Many works have attempted to reconstruct the SFH of the MW disc using a broad range of techniques on data comprised of stars in or near the solar neighbourhood, such as with classical chemical evolution modelling \citep{1989MNRAS.239..885M, Boissier1999, 2015A&A...578A..87S, 2018MNRAS.476.3432P, 2018A&A...618A..78H, 2021A&A...647A..73S}, model-driven approaches \citep{2018Natur.559..585N,2019A&A...624L...1M, 2021A&A...647A..39S, 2025A&A...697A.128D}, or colour-magnitude diagram fitting \citep{2020NatAs...4..965R, 2022MNRAS.510.4669S, 2024MNRAS.527..583M}. The key assumption of all these approaches is that the sample of the observed stars is representative of a large volume or the entire disc. SFH reconstructions of the MW based on chemical abundance data often fail to account for the survey's specific selection function. While this simplification may not significantly affect results near the solar radius, it is critical at other Galactocentric distances, where the age–abundance relations can differ substantially~\citep{2019MNRAS.489.1742F}. As a result, comparisons between models are likely to be inconsistent due to the varying mixtures of stellar populations sampled at different radii, which are heavily weighted towards the solar radius and currently provide little to no insight into other parts of the disc. 

SFH reconstructions based on Gaia photometry and CMD fitting, while relying on complete samples within specific magnitude ranges, are still subject to several limitations—even within the local 1–2 kpc region around the Sun. First, extinction significantly affects the completeness of the data; for instance, as much as 40\% of stars may be excluded from CMD fitting due to extinction effects \citep{2020NatAs...4..965R}. Second, the theoretical modelling of the giant branch remains uncertain, potentially introducing biases into the derived SFH~\citep{2024MNRAS.528.2790Z}. Lastly, the local stellar sample is known to contain numerous kinematic substructures. These may be mistakenly interpreted in CMD-based reconstructions as discrete star formation episodes. 


Additionally, stars undergo radial migration, through both blurring and churning, which causes them to drift away from their birth radii ($\rm R_{birth}$) and populate other regions of the Galactic disc. As a result, the solar neighbourhood contains stars originating from a broad range of $\rm R_{birth}$, implying that mono-age populations observed at 8 kpc do not represent the local enrichment across lookback time \citep{2023MNRAS.525.2208R}. The impact of migration on SFH has been explicitly quantified in simulations: \cite{Minchev2025} found that the net effect is to suppress and broaden SFH peaks, which can be overestimated by 100--200\% in the outer disc. \cite{Bernaldez2025} extended this to TNG50 MW/M31 analogues, linking the magnitude of the distortions to bar strength, disc thickness, and merger history. Consequently, any attempt to reconstruct the spatial evolution of the MW disc must incorporate the effects of radial migration \citep[see also][]{Francois1993, Ratcliffe2023_chemicalclocks}.

 

%

In this work, we address both issues when reconstructing the SFH of the MW. In particular, we use stellar age information from our orbit superposition calculations presented in \citealt{2025A&A...695A.220K} (hereafter \citetalias{2025A&A...695A.220K}), which corrects the observed distribution function of stellar populations to obtain the complete structural, kinematic, and chemical composition of the entire MW disc while mitigating the artifacts caused by the spatial footprint of the APOGEE DR17~(\citealt{mapping-disk, mapping-bulge}, hereafter \citetalias{mapping-disk} and \citetalias{mapping-bulge}). Secondly, we use the stars' \Rb\ instead of their current radii. The chemical homogeneity observed in stellar clusters \citep{BH2010, Ness2022} has enabled recent studies to infer the \Rb\ of stars within the MW disc \citep{2018MNRAS.481.1645M, Frankel2018, 2019ApJ...884...99F, 2020ApJ...896...15F}. Observations \citep{2019ApJ...883..177N, Sharma2022} and cosmological simulations \citep{Carrillo2023} show that stars with similar [Fe/H] and age display low dispersion in other elemental abundances, suggesting that these two properties are key tracers of a star's birth environment \citep{2022ApJ...924...60R}. Leveraging this idea, \cite{2018MNRAS.481.1645M} reconstructed stellar \Rb\ and the MW disc metallicity profile over time by requiring meaningful \Rb\ distributions. This method was refined by \cite{2024MNRAS.535..392L}, who found that the time evolution of the metallicity gradient correlates with the scatter in [Fe/H] across age. It was further improved in \citet{2025A&A...698A.267R}, who demonstrated that accounting for the temporal variation in the star-forming region is essential to accurately recover the metallicity profile of the ISM over cosmic time, improving estimates by 30\%.

In this work, we combine the recent advancements in modelling the MW disc and estimating stellar $\rm R_{birth}$ to recover the spatial and time evolution of mass growth across the Galactic disc. This paper is structured as follows. Sections \ref{sec5::data} and \ref{sec5::methods} respectively describe the data and methods used in this work. Sections \ref{sec5::results} and \ref{sec5::summary} present our results and conclusions, respectively.

\section{APOGEE data}\label{sec5::data}

In this work, we use the giant stars sample from APOGEE DR17. The details of the selection can be found in \citetalias{mapping-disk}, from which the input sample of stars is adopted. We use radial velocities, atmospheric parameters and stellar abundances~(\FeH and [Mg/Fe]) from APOGEE DR17~\citep{2022ApJS..259...35A}, which were complemented by the proper motions from \Gaia DR3~\citep{2023A&A...674A...1G}. We use stars with radial velocity uncertainties better than 2~km~s$^{-1}$, distance errors $<20\%$, and proper motion errors $<10\%$, as this is critical to calculate stellar orbits. In order to cover a larger area across the MW disc, we select giant stars with $\rm log g < 2.2$, $\rm ASPCAPFLAG=0$, and $\rm EXTRATARG=0$. Our final selection includes approximately 80,000 stars spanning the MW disc. For the analysis we adopted stellar ages from the \texttt{distmass} catalogue \citep{2024AJ....167...73S},  which have age uncertainties $<2$~Gyr. This age catalogue was extensively analysed by \cite{2023ApJ...954..124I}, who demonstrated its quality~\citep[][]{mapping-disk,2025AJ....169..280G}.

The selection of giant stars from the APOGEE catalogues, despite their stellar parameters being less precise compared to those of dwarfs, is driven by the necessity to cover a larger area of the Galaxy~\citep[see the comparison in][]{2023ApJ...954..124I}. As demonstrated in \citetalias{2025A&A...695A.220K}, this broader coverage is crucial for the effective application of the orbit superposition method. Following the same methodology as in \citetalias{mapping-disk}, we do not remove stars with high age and metallicity uncertainties from our sample as they are used to propagate information along the orbits. 

\section{Methods}\label{sec5::methods}
\subsection{Orbit superposition}\label{sec5::sos}
In this work, we use the orbit superposition results obtained in the previous papers of this series (\citetalias{mapping-disk}, \citetalias{mapping-bulge}). Here, we briefly mention the key steps of the approach, and refer readers to \citetalias{2025A&A...695A.220K} where the method is described in detail. We adopt the 3D mass distribution of the MW, including its stellar component from \cite{2022MNRAS.514L...1S}, which is an updated analytic model of the potential constructed by \cite{2017MNRAS.465.1621P}. This analytic potential, available in AGAMA~\citep{2019MNRAS.482.1525V}, shows the correct behaviour of the mass distribution outside the bar region and reproduces the 3D density of the bar, including the X-shape structure of the bulge~\citep{2013MNRAS.435.1874W, 2015MNRAS.450.4050W}.

\begin{figure}
    \centering
    \includegraphics[width=0.9\hsize]{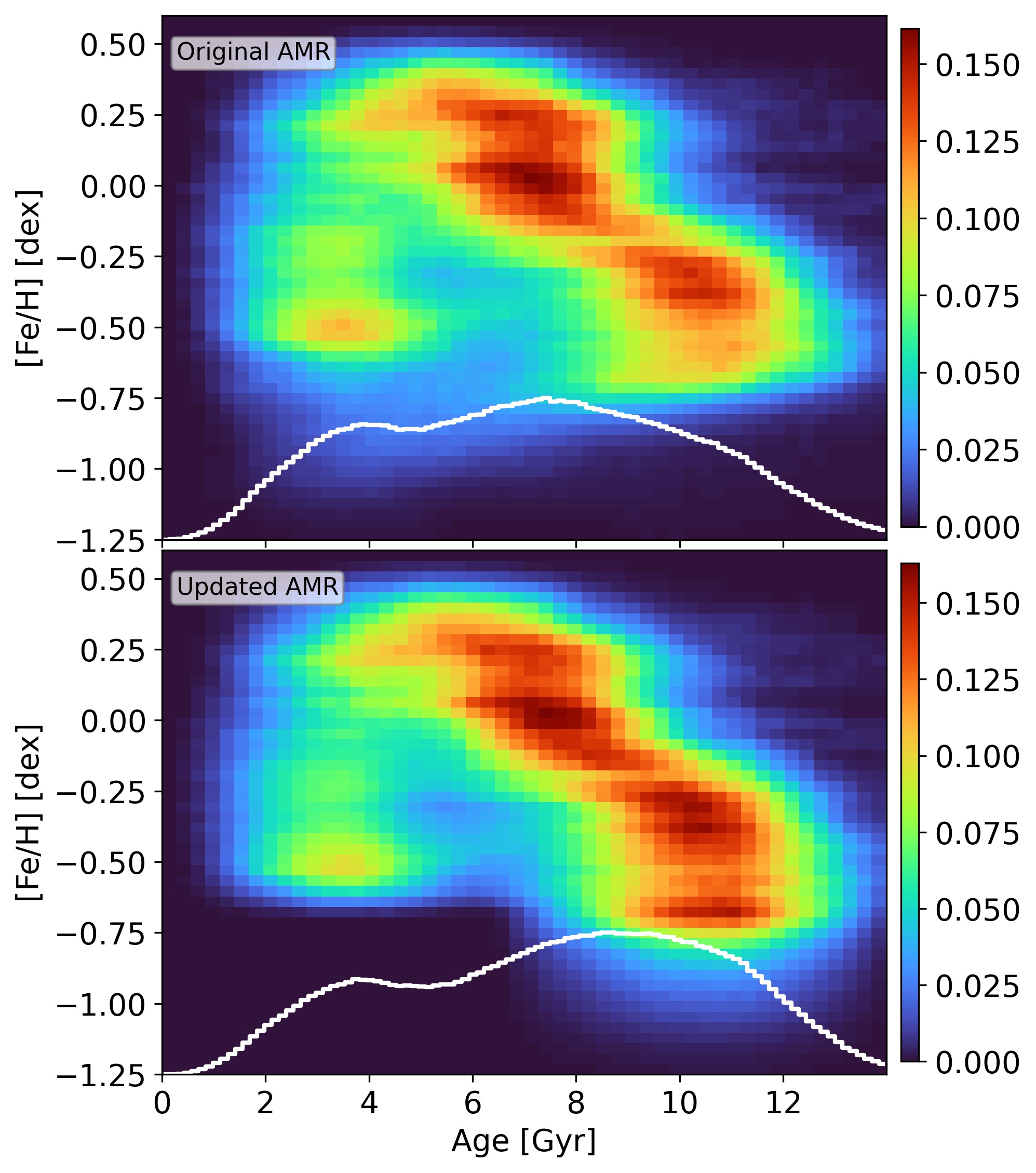}
    \caption{Age–metallicity relation of the initial stellar mass distribution derived from the orbit superposition modelling of APOGEE data. The bottom panel displays the age-metallicity relation after applying a correction to stellar ages of the young $\alpha-$rich~(YAR) and metal-poor populations. 
    These revised ages were resampled from the high-$\alpha$ population, as described in Section~\ref{sec5:fixed_ages}. The corrected YAR and metal-poor stars together account for approximately 13\% of the total stellar mass of the Galaxy.  In both panels, the white histograms trace the age mass density distribution, while the colour bar illustrates the age--metallicity density. Fig.~\ref{fig05::age_comparison} provides an additional comparison of the original and adjusted ages.
    }
    \label{fig05::AMR_update}
\end{figure}

We integrate orbits of the APOGEE stars, assuming a constant bar pattern speed of $\rm 37~km~s^{-1}~kpc^{-1}$, in agreement with various studies~\citep{2017MNRAS.465.1621P, 2019MNRAS.490.4740B, 2019MNRAS.488.4552S, 2020A&A...634L...8K, 2022MNRAS.512.2171C, 2022A&A...663A..38K}. The weights of the orbits in the rotating rest frame were calculated by adjusting their total 3D density to the analytic solution for the stellar component from \cite{2022MNRAS.514L...1S}. Each orbit was divided into 500 phase-space coordinates, and chemical abundances for stars along each orbit were sampled from normal distributions with the uncertainties as the standard deviation of the distribution. In other words, each orbit can be considered as a sample of stars following each other along the orbit with similar stellar parameters. 

The orbit superposition solution provides the present-day mass of stellar populations of different ages and metallicities across the MW. In order to account for stellar mass loss by stellar populations---specifically, winds and supernovae (SNe)---we calculated the initial mass of simple stellar populations using \texttt{ChemPy}~\citep{2017A&A...605A..59R}\footnote{\url{https://github.com/jan-rybizki/Chempy}}, assuming a single burst model for each particle, Chabrier initial mass function~\citep{2003PASP..115..763C}, and the contribution from SNeI~\citep{2013MNRAS.429.1156S}, SNeII~\citep{2013ARA&A..51..457N}, and AGB stars~\citep{2010MNRAS.403.1413K}. The stellar mass loss of a single-age population occurs quite rapidly, yielding approximately $40\%$ of the initial mass after 1--2 Gyr and a negligible amount of mass loss at later times. Throughout this study, we use the initial stellar mass of stellar populations, rather than the present-day mass distribution constrained by the orbit superposition model. 

\subsection{Adjustment of ages of stellar populations}\label{sec5:fixed_ages}
To analyse the stellar mass growth as a function of lookback time, two key conditions must be met: first, the spatial coverage of the sample must be representative of the entire Galaxy and second, stellar ages must be sufficiently precise. The first requirement is addressed by our orbit superposition method, which ensures spatial completeness. However, the reliability of age estimates lies beyond our direct control. In practice, stars with uncertain ages are often excluded from analyses, potentially introducing subtle and non-trivial biases. In this study, such simplifications are not applicable and are addressed differently.

A fraction of the stars in our sample have improper ages as they are members of the young $\alpha-$rich (YAR; high-$\alpha$ stars with ages $<7$ Gyr; see e.g. \citealt{2015MNRAS.451.2230M,2023A&A...676A.108C}) or metal-poor populations with poor convergence of the age determination~($\FeH<-0.65$~dex; \citealt{2024AJ....167...73S}). Keeping these stellar ages in our analysis would make the MW's mass build-up slower than expected, as some mass during early disc formation would appear spread across a wider time frame. It is assumed these populations have artificially lower ages due to limitations of the age catalogue (\citealt{2023ApJ...954..124I}) and mass transfer \citep{2023A&A...671A..21J}; we follow \cite{Imig2025} and assign them older ages in line with predictions from chemical evolution models and observations \citep{2006ApJ...653.1145K, 2021A&A...647A..73S}. Since the metal-poor stars with uncertain ages represent the high-$\alpha$ populations, we resampled their ages using more reliable samples from the high-$\alpha$ sequence. For the YAR sample, we followed a similar procedure and drew new ages from high-$\alpha$ stars in $0.1$~dex [Fe/H] abundance bins. This approach is further justified because these YAR stars resemble genuine high-$\alpha$ members in other properties, including their kinematics \citep{2023A&A...676A.108C}. The comparison of the updated age--metallicity relation of our sample with the original one is provided in Fig. \ref{fig05::AMR_update}. While the original age--metallicity relation (top panel) shows evident inconsistencies at low metallicities, our adjusted version appears more physically plausible and aligns well with the expectations from independent age catalogues with robust age estimates \citep[e.g.,][]{2022Natur.603..599X,2024A&A...692A.243C}. Overall, the age adjustments affect about $13\%$ of the total stellar mass. The effect of reassigning artificially young stars to older populations within the high-$\alpha$ sequence is clearly illustrated by the white histograms in Fig.~\ref{fig05::AMR_update}, which indicate a slight modification of the age distribution without significantly altering its overall structure. While our approach is motivated by clear shortcomings in current age determinations, it does not fully resolve these issues. There remains room for improvement, which will be enabled by the availability of more precise age catalogues from future data releases. 

\subsection{Stellar birth radii estimation}\label{sec5:rbirth}

\begin{figure}
    \centering
    \includegraphics[width=.87\hsize]{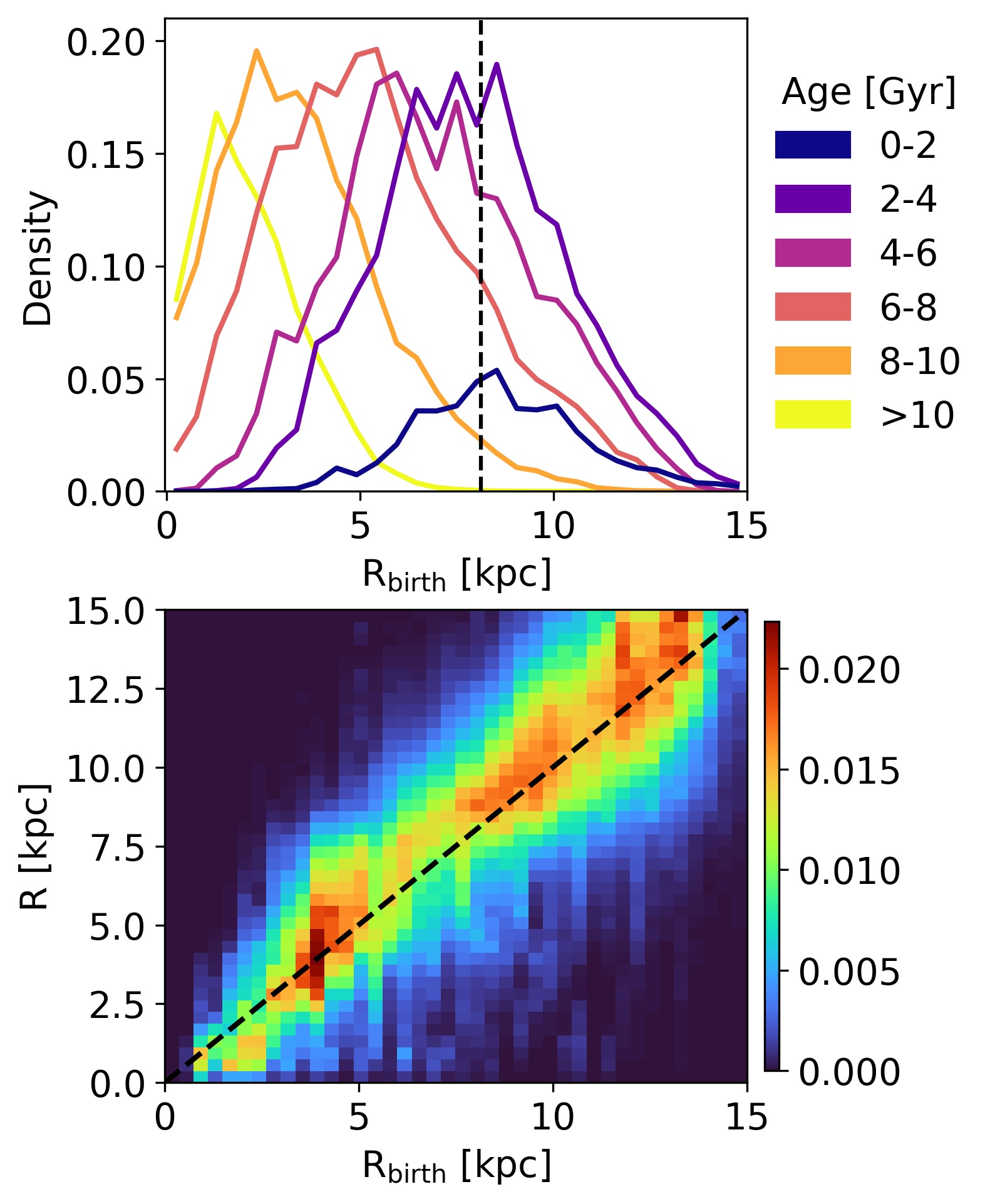} 
    \caption{Verification of stellar \Rb. {\it Top:}  Distribution of \Rb\ of stars in mono-age populations currently located in the solar neighbourhood~($\rm |z| < 1.5$ kpc; $\rm |R-8.125|<0.15$~kpc) with eccentricity $<0.5$. Older stars in the solar neighbourhood originate in the inner disc while younger populations are born locally, in agreement with expectations. {\it Bottom:} Stellar mass distribution in the R--\Rb\ (i.e., current--birth) plane for stars with age $<4$ Gyr, showing overall minimal net migration in the most recent Gyrs.}
    \label{fig05::Rb_method}
\end{figure}

To estimate stellar \Rb, we used the publicly available package \texttt{Rbirth}\footnote{\url{https://github.com/BridgetRatcliffe/Rbirth/}}, which implements the method of \cite{2025A&A...698A.267R}, improving upon the techniques described in \cite{2024MNRAS.535..392L,2023MNRAS.525.2208R}. This method assumes a linear radial metallicity gradient in the ISM ($\rm \nabla[Fe/H](\tau)$), which is a reasonable assumption for the Galactic disc supported by both observations \citep{2017MNRAS.471..987E, 2021MNRAS.502..225A} and simulations \citep{2018MNRAS.478..155V, 2022MNRAS.515L..34L} with any second order term being small \citep{2025OJAp....8E..47B}. 

As shown in \cite{2024A&A...690A.352R}, recovering $\rm \nabla[Fe/H](\tau)$ from the scatter in [Fe/H] across age works best for galaxies with bar strengths similar to the MW. However, even in this regime, the recovered gradient is too strong in recent Gyrs and too weak at larger lookback times due to growth of the star-forming region. Thus, we use the correction provided in \cite{2025A&A...698A.267R} that accounts for the time variation in the width of the star-forming region. The present day value for $\rm \nabla[Fe/H](\tau)$ was chosen to be -0.064 dex/kpc \citep{2024A&A...690A.246T} and the steepest gradient was chosen as the value that minimized $\rm |R \text{--}R_{birth}|$ for the youngest stars ($-0.151$ dex/kpc). 

The obtained $\rm R_{birth}$ distributions of solar neighbourhood stars are provided in the top panel of Fig. \ref{fig05::Rb_method}. To isolate in situ disc stars, we restrict the sample to stars with eccentricities below 0.5. As expected \citep[e.g.,][]{2021MNRAS.503.5826A}, the majority of old stars near the Sun were born in the inner Galaxy, whereas younger stars predominantly formed near their current location. Although the figure indicates that outward migration is the dominant trend, we also observe a modest population of stars that have migrated inward. In order to verify the \Rb\ estimates on a larger scale, we also provide R vs \Rb\ for young stars (age $<4$ Gyr) in the bottom panel of Fig.~\ref{fig05::Rb_method}. There is a near one-to-one trend with a weak preference towards outward migration, enhanced near the bar end~(4--5~kpc), in agreement with expectations \citep{2014A&A...572A..92M, 2021MNRAS.506..759V, 2024A&A...690A.147H}. This suggests that despite the presence of a bar and spiral arms in the outer disc, the net migration is limited over the last $4$ Gyr. 

\begin{figure*}
    \centering
    \includegraphics[width=1\hsize]{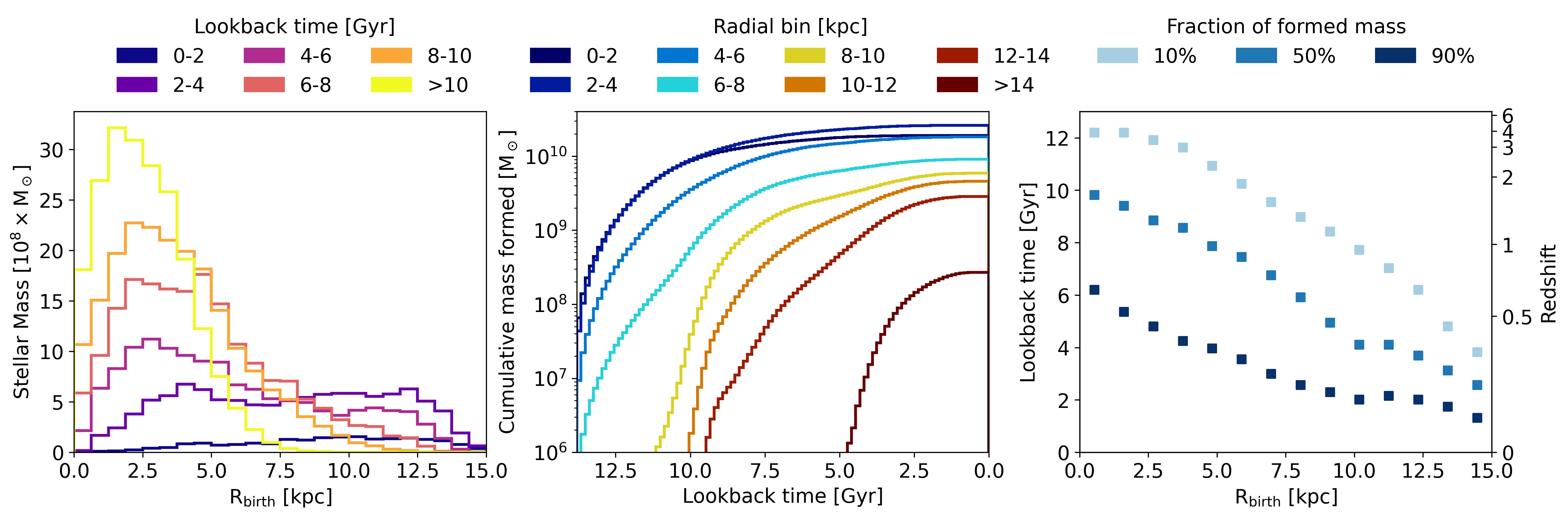}
    \caption{Inside-out mass growth of the MW stellar disc. {\it Left:} \Rb\ distribution of stellar mass in different bins of the lookback time~(stellar age). 
    {\it Middle:} Cumulative mass formed at a given \Rb\ over time. The inner disc (blue lines) exhibits an initial growth with 80\% of its mass created by 6--7 Gyr ago, while the outer disc (red lines) formed 80\% of its mass $<4$ Gyr ago.
    {\it Right:} Lookback time distribution of stellar mass formation within a 1-kpc-wide radial bin, expressed as a percentage of the total mass formed at the corresponding radius. The symbols show at what lookback time $10\%$~(light blue), $50\%$~(blue) and $90\%$~(dark blue) of mass formed at a given distance from the centre.   }
    \label{fig05::stellarMass}
\end{figure*}

\section{Results}\label{sec5::results}
Before presenting our results, we highlight the key components of our methodology that enable us to trace the temporal evolution of the Galaxy:
\begin{itemize}
\item We use initial stellar masses, corrected for mass loss, rather than present-day mass distributions (Section~\ref{sec5::sos}).
\item Ages for the YAR and metal-poor stars are resampled to improve the age reliability for these populations. This ensures a consistent coupling between chemical abundances, ages, and completeness of stellar mass across the MW~(Section~\ref{sec5:fixed_ages}).
\item The evolutionary history of the MW is reconstructed based on the estimated \Rb\ of stellar populations~(with precision of 1.5 kpc), enabling the analysis of both star formation and the subsequent redistribution of stars over time~(Section~\ref{sec5:rbirth}).
\end{itemize}

\subsection{Stellar mass growth across the MW disc}\label{sec5::results_mass}

We find that early star formation was confined to the inner $5$ kpc for the first $\sim 3$ Gyr of the MW disc's evolution, as shown in the left panel of Fig. \ref{fig05::stellarMass}, which provides the stellar mass formed in the MW disc as a function of $\rm R_{birth}$ for different lookback times. Assuming minimal outward migration at early times (lookback time $>8$ Gyr, or $z>1.5$), we find that the effective radius ($\rm R_{\rm eff}$)---the radius enclosing half the stellar mass---of a $\sim$4--6 Gyr-old MW was $\sim$2 kpc, consistent with star-forming galaxies at similar epochs \citep{2014ApJ...788...28V, 2015ApJS..219...15S}. It is crucial to emphasise that only a very small fraction of stars were able to form beyond $8$~kpc at this time. As a result, the solar radius was not initially populated by in-situ star formation, but could have been reached by stars that had migrated outward from inner regions. This has significant implications for models of Galactic chemical evolution that aim to reproduce the abundance patterns observed near the Sun. In particular, these models should account for the fact that in-situ star formation at the solar radius likely began only around 8--10~Gyr ago and proceeded at a relatively modest rate.

While star formation is relatively confined to $\rm R_{birth} < 8$ kpc for lookback times $>6$ Gyr ago with large amounts of mass being produced (65\% of the total disc mass), the remaining parts of the Galactic disc started to form rapidly. Younger stellar populations exhibit progressively broader distributions in Galactocentric radius, consistent with an increasing scale length of the SFR over time. 

Despite this global trend, there is still a prominent peak at $\rm R_{birth} \approx4$ kpc for lookback times $<4$ Gyr, in addition to a peak 6--8 Gyr ago at \Rb\ $\approx$ 5 kpc. Given the MW's bar length is believed to be 3.5--5 kpc \citep{2017MNRAS.465.1621P, 2015MNRAS.450.4050W, 2019MNRAS.490.4740B, 2023MNRAS.520.4779L, 2024MNRAS.533.3395Z}, the peak in mass at $\rm R_{birth} = $ 4--5 kpc may be indicative of the star formation often peaking at the bar ends or along bar–spiral connections, which is also found in external galaxies \citep{2004A&A...414...23J,2007A&A...474...43V, 2018MNRAS.474.3101J, 2020MNRAS.499.1116F}. In the MW, the presence of a star-forming ring near 5 kpc \citep{1988ApJ...327..139C} remains uncertain~\citep{2012MNRAS.421.2940D, 2014MNRAS.444..919P, 2016MNRAS.455.1782K}. However, the tightly wound spiral arms, which appear to connect to the ends of the bar, may enhance the local gas density and thereby trigger star formation~\citep{2025arXiv250202651M}. This effect is observed in Fig.~\ref{fig05::stellarMass}, where an increase in stellar mass formation is seen at radii of approximately 4--5 kpc.

\begin{figure*}
    \centering
    \sidecaption
    \adjincludegraphics[width=0.65\linewidth]{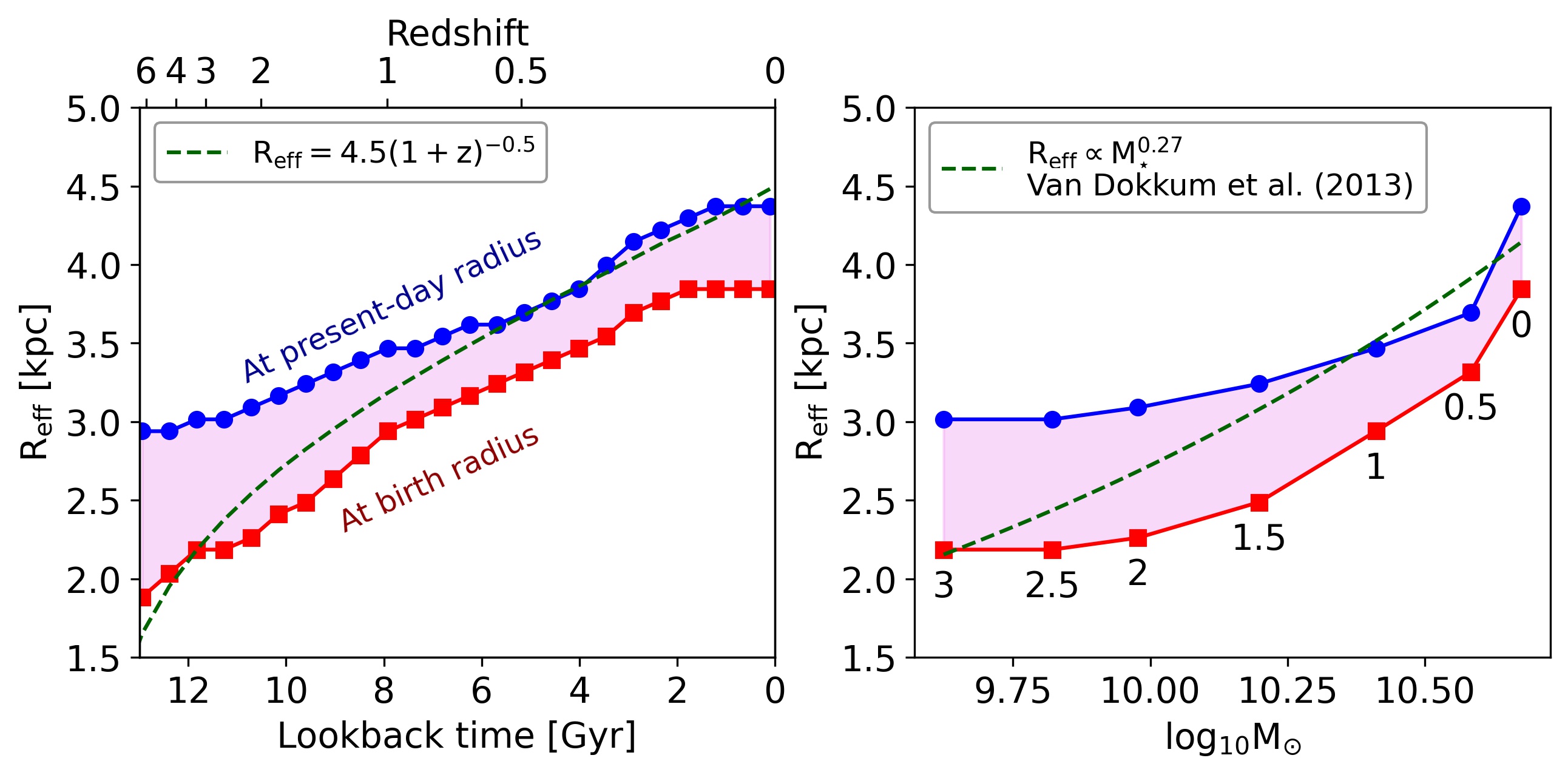} 
    \caption{Evolution of the MW's $\rm R_{eff}$.
    {\it Left:} Time evolution of the stellar $\rm R_{eff}$ under two limiting assumptions: stars remain at their \Rb\ (red squares), or stars are placed at their present-day radii immediately after formation (circles). 
    {\it Right:} $\rm R_{eff}$--stellar mass relation, where stellar mass reflects time-dependent mass loss, measured every 0.5 redshift as shown in black text. The true evolution of $\rm R_{eff}$ is expected to lie between these two regimes, as illustrated by the shaded region in both panels, and should be close to the red lines at early epochs, with a tendency towards the blue lines at the present day. In both panels, the green dashed lines are not formal fits but are included as a reference, including from \cite{2013ApJ...771L..35V}.
    }
    \label{fig05::Reff}
\end{figure*}

\begin{figure}
    \centering
    \includegraphics[width=\hsize]{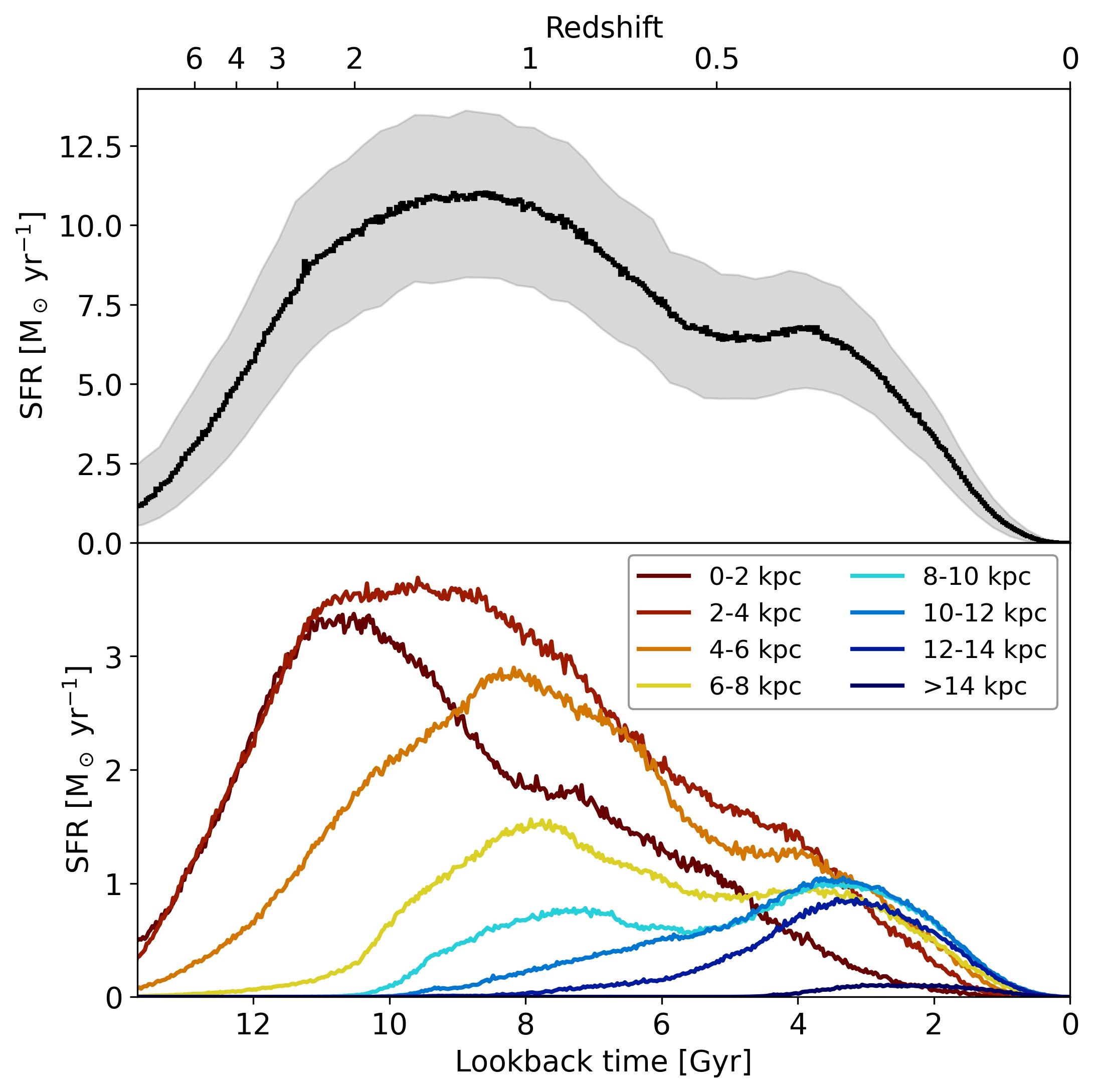}
    \caption{SFH of the MW disc. {\it Top:} Stellar mass formed per unit time as a function of lookback time, normalised by the time bin width. The shaded areas show the 16--84\% confidence of a given SFH. {\it Bottom:} Evolution of the SFR in bins of stellar \Rb. The SFR at different \Rb\ per unit area is provided in Fig. \ref{fig05::sfh_area}. 
    }
    \label{fig05::SFR_total_radial}
\end{figure}

\begin{figure*}
    \centering
    \includegraphics[width=\hsize]{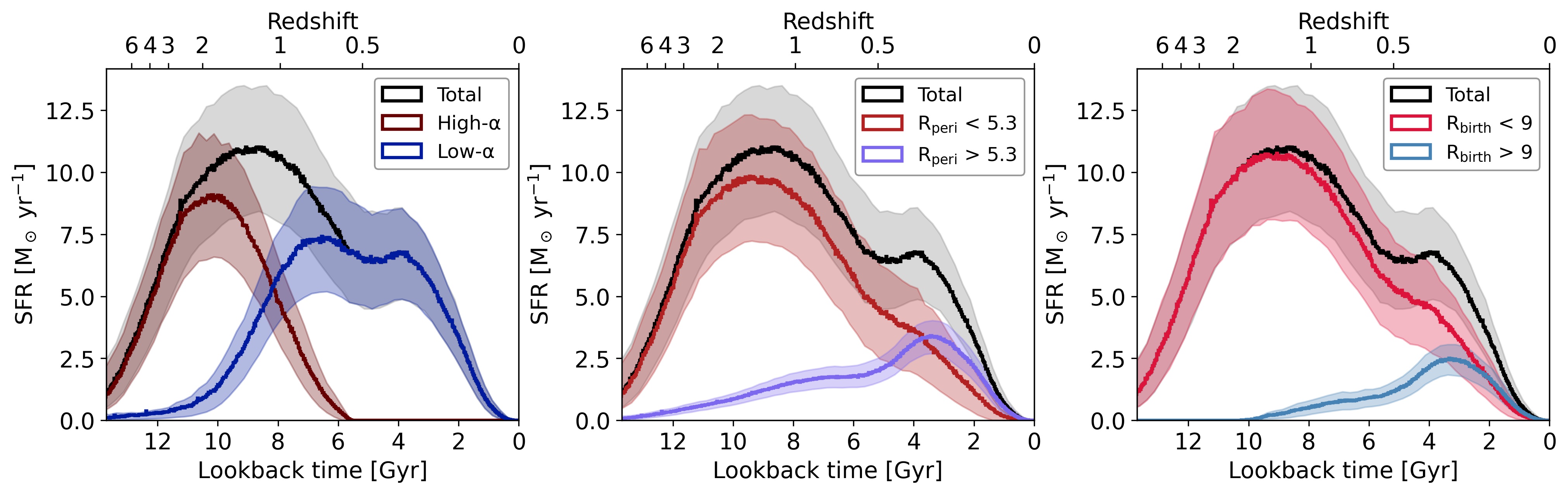}
    \caption{SFH of the MW disc and its sub-populations. The black curve represents the stellar mass formed per unit time as a function of lookback time, normalised by the time bin width.
     {\it Left:} The breakdown of the SFR into high- and low-$\alpha$ populations, in red and blue, respectively.
    {\it Middle:} The SFRs of the present-day kinematically-defined inner ($\rm R_{peri} < 5.3$ kpc, red) and outer ($\rm R_{peri} > 5.3$ kpc, violet) discs.
    {\it Right:}  The SFRs of populations formed inside and outside of 9 kpc, in red and blue, respectively.
    In each panel, the shaded areas show the 16--84\% confidence of a given SFH.}
    \label{fig05::SFR_total_different_selections}
\end{figure*}

The middle and right panels of Fig. \ref{fig05::stellarMass} provide the cumulative mass evolution and the onset and termination of star formation---marked by 10\% and 90\% mass fractions respectively---for different regions across the Galactic disc. The inner disc (blue lines) formed its stars early on, with 50\% and 90\% of its mass created by a lookback time of 8--10 Gyr ago and 5--6 Gyr ago, respectively (in agreement with \citealt{2014ApJ...781L..31S}). This is in agreement with \cite{2013ApJ...766...15P}, who found that the inner disc had most of its mass growth by redshift 2 in external galaxies. On the other hand, the outer disc radial bins (red lines) formed stars much later, where the cumulative mass went from 50\% to 90\% within $<2$ Gyr. This more rapid increase occurs after a slower growth phase until approximately 4--6~Gyr ago. Given this timing, the feature may correspond to the first infall of the Sagittarius dwarf galaxy \citep{1994Natur.370..194I}. While we find the outer disc is predominantly affected at this time (in agreement with the effects of a merger; \citealt{2024MNRAS.527.2426A}), \citetalias{mapping-disk} discussed that the direct gas contribution from the dwarf galaxy may be insufficient to support the formation of the required stellar mass, whose dynamical influence at first approach was 30--50~kpc ~\citep{2018MNRAS.481..286L}. If such an external perturber were responsible, one would expect an outside-in triggering of star formation~\citep[e.g.][]{2018MNRAS.479.3381B}, which is not apparent in Fig. \ref{fig05::stellarMass}.


The right panel of Fig.~\ref{fig05::stellarMass} illustrates that star formation begins later at larger Galactocentric distances, while in the inner disc it ceases earlier, indicating an inside-out quenching pattern. This behaviour is likely driven by the gradual exhaustion of the gas reservoir commonly referred to as strangulation~\citep[e.g.][]{1980ApJ...237..692L,2015Natur.521..192P}, as well as by bar-driven suppression or, more generally, morphological quenching of star formation~\citep{2009ApJ...707..250M, 2020A&A...644A..79G, 2024A&A...687A.255S}. Notably, the stellar mass formation timescale decreases steadily out to approximately 10 kpc, beyond which it flattens, suggesting that the outer disc formed over a relatively short timescale.

To quantify the build-up rate of the MW disc, Fig.~\ref{fig05::Reff} presents the evolution of its stellar $\rm R_{eff}$ as a function of lookback time~(left panel) and total stellar mass~(right panel). The red squares show the case where stars remain at their \Rb, i.e. no radial migration. In contrast, the blue circles represent the opposite extreme: stars are placed at their present-day radii immediately after formation. While both assumptions are unrealistic, they serve respectively as lower and upper boundaries on the $\rm R_{eff}$ evolution. The true evolution of the MW’s $\rm R_{eff}$ lies between these two extreme scenarios~(pink area), resembling the no migration case (red curve) at early times, corresponding to lower stellar masses in the right panel, and gradually approaching the instantaneous migration case (blue curve) toward the present day. In both cases, $\rm R_{eff}$ is computed from the total stellar mass formed up to the corresponding time, accounting for mass loss as well.

Our calculations suggest that the MW's  $\rm R_{eff}$ increased from $\approx 2$~kpc in the early stages~($z \sim 2\text{--}3$) to $\approx 4.3$~kpc in the present-day. These boundary values are consistent with both extragalactic observations~\citep[e.g.][]{2013ApJ...771L..35V, 2014ApJ...788...28V, 2015ApJS..219...15S, 2019ApJ...880...57M, 2020ApJ...905..170M} and independent estimates of the current stellar half-mass radius of the MW~\citep{2016ARA&A..54..529B} and nearby galaxies of similar mass~\citep{2015MNRAS.447.2603L, 2022MNRAS.509.3751H, 2020MNRAS.493...87T}. While the growth is broadly monotonic, an episode of more rapid disc ``expansion'' begins around 4 Gyr ago (z $\sim0.5$)~(Fig.~\ref{fig05::Reff}). This is particularly noteworthy given that relatively little stellar mass has formed since that time, especially in the inner Galaxy. Thus, the increase in $\rm R_{eff}$ primarily reflects the outward extension of the outer disc. 

While $\rm R_{eff}$ is known to have limitations as a metric for galaxy size~\citep{2020MNRAS.493...87T,2022A&A...667A..87C}, its widespread use in the literature justifies its adoption here to assess the impact of radial migration on the evolution of disc size, as shown in Fig.~\ref{fig05::Reff}. For completeness, the evolution of the disc scale length ($\rm h_d$) is also presented in the Appendix (Fig. \ref{fig05::hd}). Radial migration appears to increase the disc size by $\approx 0.5$~kpc. While this absolute change may seem modest, it represents roughly a 20\% contribution to the total disc size, underscoring the significance of outward stellar migration in shaping the present-day structure of the MW. However, the impact of radial migration is expected to be significantly greater at specific Galactocentric radii~\citep{2010ApJ...722..112M}, particularly in the outer disc~\citep{2008ApJ...675L..65R}, where migration is strongly influenced by resonant interactions. This aspect is explored in more detail in the following sections.

\subsection{Spatially resolved SFH of MW}\label{sec5::results_SFH}

In the previous section, we focused on the build-up of stellar mass in the Galactic disc. Now, we discuss the SFH of the MW disc and its components, including also its radial variations.
We emphasise that the SFH presented here offers a global, somewhat smoothed representation, which does not capture short-timescale variations or sharp features. This is an intentional consequence of our methodology, which relies on stellar ages derived from spectroscopic data with a typical uncertainty of $\approx 1.5$~Gyr~\citep{2024AJ....167...73S}, a factor that is explicitly incorporated into our solution~(see Section~\ref{sec5::sos}) together with a smoothing of old ages due to re-sampling of YAR and metal-poor populations~(see Section~\ref{sec5:fixed_ages}) naturally leading to a smoothing of the inferred SFH. This approach is motivated by the desire to avoid over-interpreting small-scale fluctuations that are not robustly constrained by current data and/or methodology. Instead, our aim is to provide a broad and reliable overview of the MW's SFH. Future stellar age catalogues, offering improved precision and homogeneity, will enable a more detailed reconstruction and allow for a more nuanced analysis of the Galaxy's temporal evolution.

The black line in the top panel of Fig. \ref{fig05::SFR_total_radial} shows the time evolution of the SFR of the entire MW disc obtained by taking into account mass loss as a function of time of the present-day stellar mass distribution obtained from the orbit superposition solution~(see Section \ref{sec5::sos}). The total SFR of the MW disc rises rapidly in the early evolutionary stages, reaching its peak approximately $9$ Gyr ago (z $\sim1.5$). Thereafter, it exhibits an overall decline toward the present day. Notably, the timing of the peak SFR aligns closely with the global cosmic SFH which shows a maximum around redshift $z\sim 2$~\citep[corresponding to 9--10 Gyr ago;][]{1998ApJ...498..106M, 2014ARA&A..52..415M}. This suggests that, like most galaxies in the universe, the MW underwent a phase of rapid halo growth accompanied by a high cold gas fraction, likely sustained by efficient gas accretion within a relatively dense cosmic environment amid frequent mergers. During this period, star formation had not yet been significantly suppressed by feedback processes or gas depletion. The SFH we recover, with a peak near $z\sim 1.5$ (9 Gyr ago), is broadly consistent with the shape of the cosmic SFH but appears somewhat delayed~($\approx 2$ Gyr) at high redshift when compared to measurements by e.g., \citet{2015A&A...578A..87S}, which place the peak around $z\sim 2\text{--}3$ (10--12 Gyr ago). Those works imply early rapid star formation in the MW, when contrasted with MW/M31 analogues in TNG50, where the MW analogues on average acquire their mass slightly slower~\citep{2024MNRAS.533.3975K}, which aligns better with our results presented here.

The MW's SFR evolution deviates from a monotonic decline since $z=1.5$. A secondary enhancement in star formation is evident, peaking around $4$ Gyr ago, suggesting a more complex evolutionary path possibly linked to late gas accretion. As discussed in Section \ref{sec5::results_mass}, this is the same time as the onset of the accelerated outer disc formation~(see also \citetalias{mapping-disk}) and steepening in the metallicity gradient \citep{2025A&A...698A.267R}. 

Now we move from discussing the disc as a whole to focusing on the spatially resolved mass formation over time. The bottom panel of Fig. \ref{fig05::SFR_total_radial} shows the SFR of the MW disc at different regions of the Galaxy using the \Rb\ of our mass-weighted sample. The SFH clearly shows that each region of the Galaxy had its own rather specific SFH, in addition to the MW exhibiting signs of inside-out formation, where star formation begins in the inner disc and progressively moves outward with cosmic time. The MW disc forms with rapid star formation within $\rm R_{birth} < 4$ kpc until a lookback time of 10 Gyr. At this time, star formation in the central region begins to slow down while it continues to increase in the $\rm 2 < R_{birth} < 6$ kpc region. About 10 Gyr ago, the MW disc began forming stars between $\rm 6 < R_{birth} < 10$ kpc, with a peak in star formation for most of the radial bins 8 Gyr ago. After this burst, the inner 4 kpc saw a reduction in its star formation until the present day. Conversely, $\rm R_{birth} > 4$ kpc saw another star formation boost about 4 Gyr ago before forming fewer stars. 

The decomposition of the SFH by stellar \Rb\ reveals the origin of the bimodal SFH pattern~(Fig.~\ref{fig05::SFR_total_radial}): the inner disc forms over an extended period, contributing the bulk of the stellar mass, while the outer disc emerges more recently, exhibiting a shorter and relatively less intense phase of star formation. This spatially resolved view highlights a clear inside-out growth of the disc, where early, sustained star formation in the inner regions, while decreasing over time, is followed by the delayed and lower-efficiency build-up of the outer disc. In this context, the precise distinction between the inner and outer discs remains somewhat ambiguous, particularly when considering present-day stellar populations. To address this and to connect the total SFH with commonly adopted definitions of stellar components, Fig.~\ref{fig05::SFR_total_different_selections} decomposes the total SFH into two components using three different criteria: chemical~(left panel), kinematic~(middle panel), and \Rb-based~(right panel) selections. This comparison provides a basis for relating the global SFH to specific stellar populations and formation channels.

The left panel of Fig.~\ref{fig05::SFR_total_different_selections} presents the SFHs of the high- and low-$\alpha$ populations, defined in Fig. 3 of \citetalias{mapping-disk} (see also Fig. \ref{fig05::alpha_sequence_def}). Notably, the high-$\alpha$ SFH is relatively brief and exhibits a distinct peak at $z \approx 2$, in close agreement with the cosmic SFH. In contrast, the low-$\alpha$ population forms over a more extended period and displays two prominent components, peaking at approximately 7 and 4~Gyr ago, with the temporal transition between the high- and low-$\alpha$ sequences overlapping~(see details in Sec.~\ref{sec5::discussion_alpha}). As previously demonstrated, the 4~Gyr component corresponds to the outer disc, or more precisely, the metal-poor ($\FeH < 0$~dex) tail of the low-$\alpha$ population. The $\sim 7$ Gyr component, on the other hand, represents the dominant contribution to the low-$\alpha$ sequence and is associated with the super-solar to extremely metal-rich stars~\citep{2024ApJ...975..293R} concentrated in the innermost regions of the MW.

The middle panel of Fig. \ref{fig05::SFR_total_different_selections} presents the SFH of the inner and outer disc when the disc is divided based on the present-day pericenter radius ($\rm R_{peri}$), hence relying on the kinematics of stellar populations. This figure makes it clear that the peak 4 Gyr ago is predominantly due to an enhancement in SFR in the outer disc, while the inner disc is relatively unaffected. In \citetalias{mapping-disk} we showed that the stars with $\rm R_{peri}$ $>5.3$ kpc are members of the low-$\alpha$ sequence with [Fe/H] $<0$, while the high-$\alpha$ sequence and the metal-rich end of the low-$\alpha$ sequence have pericenters $<5.3$ kpc. This indicates that the low-$\alpha$ sequence formed with two peaks in its SFH; the metal-rich end formed first, followed by a boost that formed the metal-poor population. 

A qualitatively similar picture is presented when decomposing the MW's SFH into inner and outer disc components defined by a \Rb\ threshold of $
9$~kpc (right panel of Fig. \ref{fig05::SFR_total_different_selections}).
It is important to note that the kinematic and \Rb–based selections of MW stellar populations rely on fundamentally different and non-overlapping sets of parameters. The kinematic classification is based on the present-day orbital properties, derived from full 6D phase-space information in combination with an assumed Galactic potential. In contrast, the \Rb\ classification is inferred solely from stellar ages and chemical abundances, independent of the current orbital configuration and any kinematic information.

\subsection{Formation of the $\alpha$-bimodality}
\label{sec5::discussion_alpha}

\begin{figure}
\centering
   \includegraphics[width=0.95\hsize]{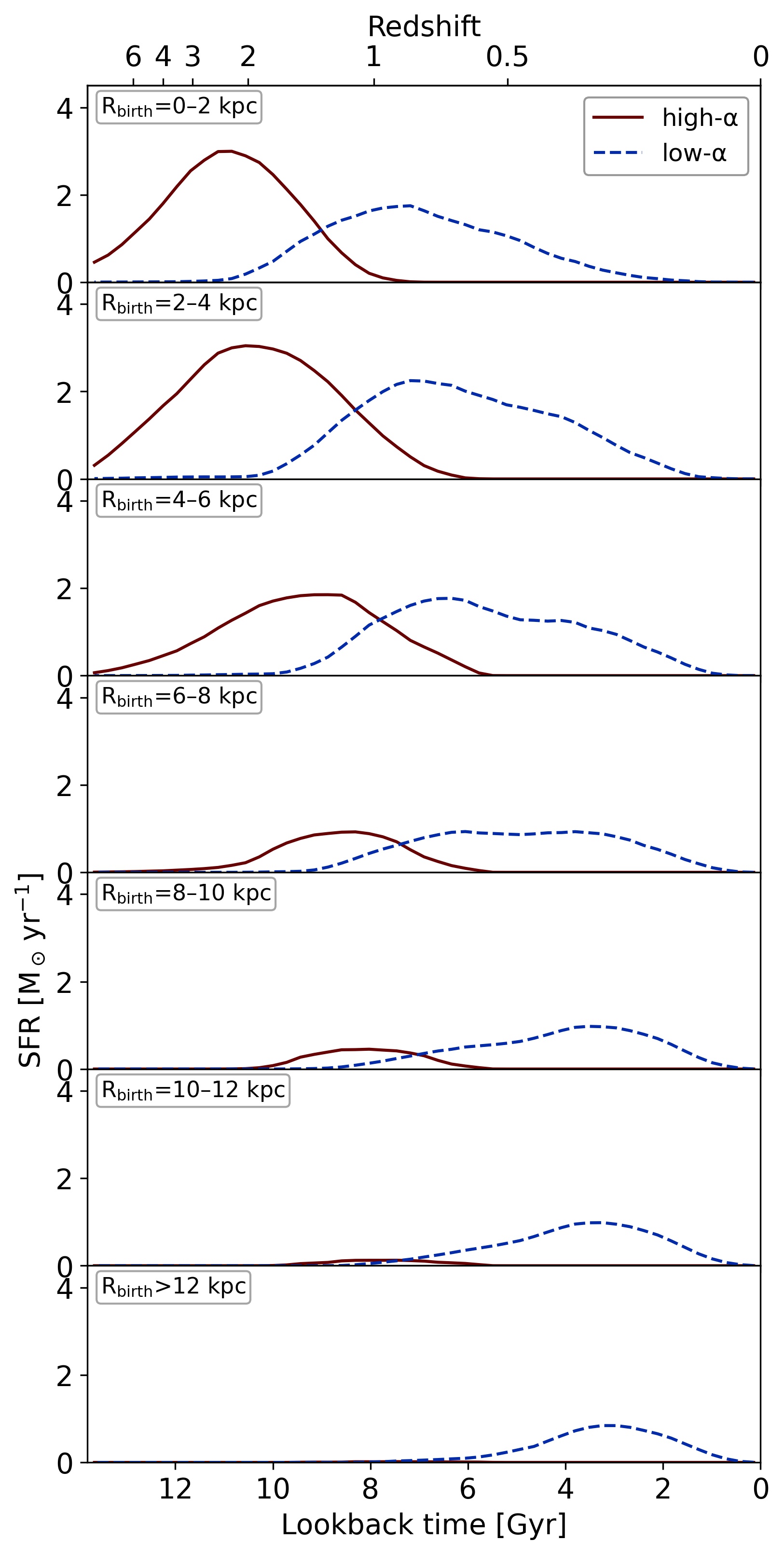}
    \caption{
    Spatial SFR of the high- (solid lines) and low-$\alpha$ (dashed lines) sequences. The high-$\alpha$ sequence formed in the inner disc (<10~kpc) and its bulk SFR moved outwards until finishing $6$ Gyr ago. The low-$\alpha$ sequence began forming in the inner disc 8--10 Gyr ago and moved outwards with time. The overlap at a given radius and time between the sequences stems from the age uncertainty (see Fig. \ref{fig05::alpha_ageUncertainty}.).
    }
    \label{fig05::sfr_alphaseqs}
\end{figure}

\begin{figure*}
    \sidecaption
    \centering
    \adjincludegraphics[width=0.68\linewidth,clip,trim={{.315\width} 0 0 0}]{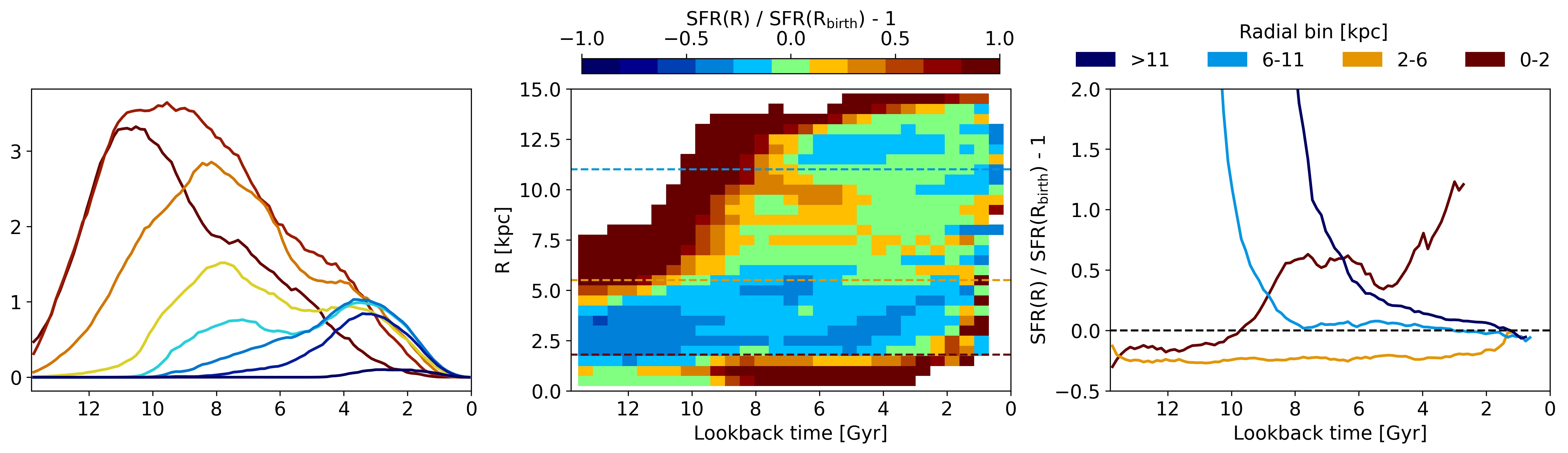} 
    \includegraphics[width=0.31\hsize]{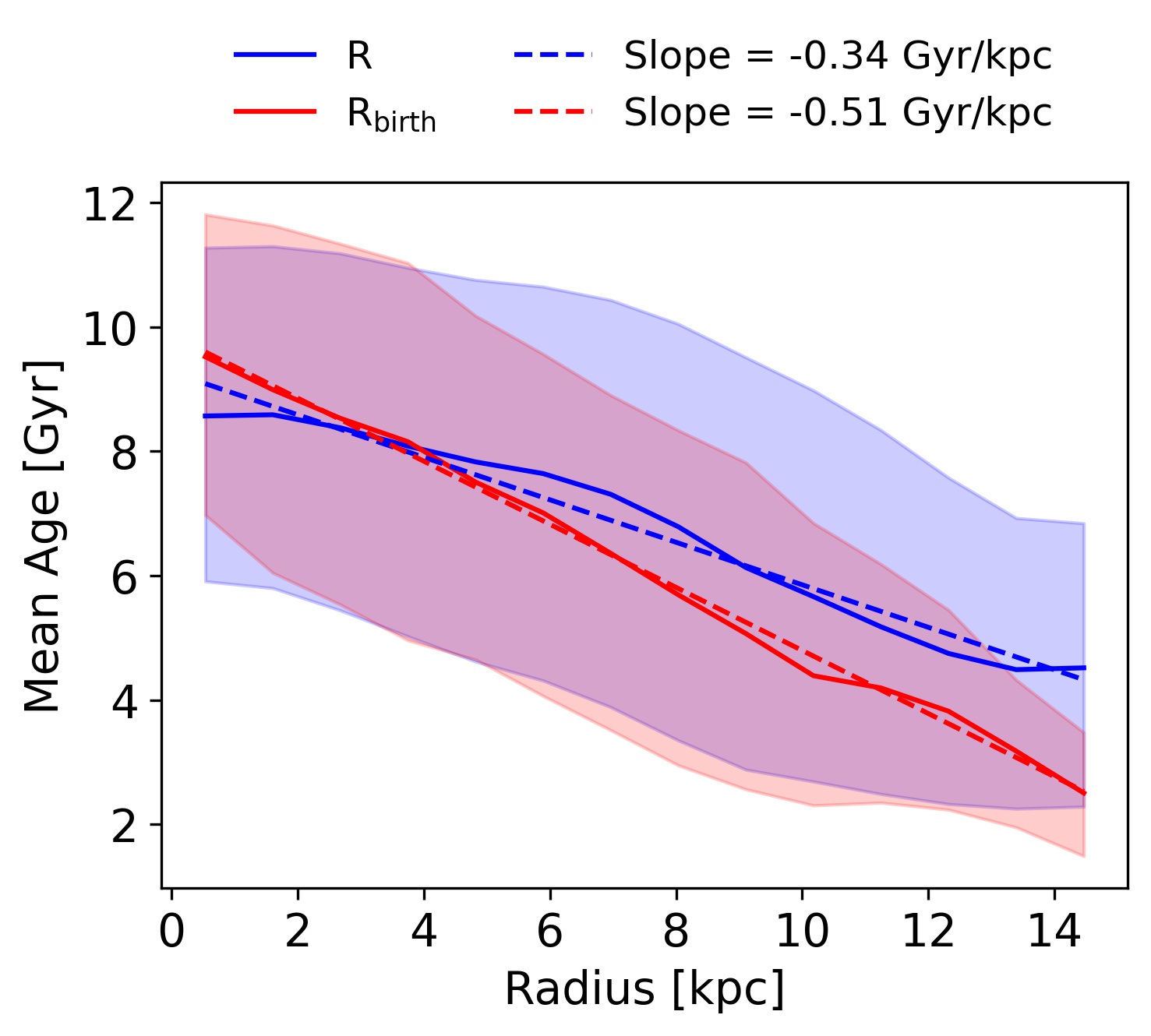}       
    \caption{
    Effect of stellar radial migration on the SFH across the MW disc.
    {\it Left and Middle:} Fractional difference between the true SFR and the present-day age distributions of mono-radius populations, shown as a function of lookback time. Positive values indicate an overestimate of SFR: more stars are currently observed at a given radius than were actually formed there at that epoch, due to inward or outward migration. The horizontal dashed lines in the left panel mark the boundaries of radial regions shown in the middle panel.
    {\it Right:} Mean age present-day~(blue line) and by-formation~(red line) profiles together with the 16--84th percentiles highlighted with the filled areas of the same colours. The mean metallicity profiles are given in Fig.~\ref{fig05::feh_gradient}.} 
    \label{fig05::SFR_migration}
\end{figure*}

The origin of the $\alpha$-dichotomy in the MW~(bimodal distribution of $\rm [\alpha/Fe]$ at \FeH< 0~dex) remains one of the most debated questions in Galactic archaeology, in part due to its degeneracy and complexity of chemical abundance patterns across the Galaxy. Previous works have shown that the $\alpha$-bimodality can form from a variety of mechanisms, including sequential evolution from gas accretion \citep{2018MNRAS.477.5072M, 2019A&A...623A..60S, 2020MNRAS.491.5435B, 2020MNRAS.497.2371L}, a titling disc \citep{2021MNRAS.503.5868R}, a natural product of secular evolution independent of merger history \citep{2021MNRAS.501.5176K, 2023MNRAS.523.2126P}, and co-formation in different physical environments~\citep{2019MNRAS.484.3476C, 2021MNRAS.502..260B}. We stress that the discussion regarding high- and low-$\alpha$ populations is independent of the geometric discs \citep{2019MNRAS.486.1167B}.

With \Rb\ estimates of our mass-weighted sample, we can get a more comprehensive picture of the MW disc and how different regions evolved. Fig. \ref{fig05::SFR_total_radial} revealed that there was no global hiatus in star formation, which was also shown to hold for individual radial bins. As shown in Fig. \ref{fig05::sfr_alphaseqs}, the high- and low-$\alpha$ sequences formed simultaneously between 10 and 6 Gyr ago, with the high-$\alpha$ sequence peaking at slightly larger radii during this time. Therefore, if a negative metallicity gradient was already in place during the formation of the high-$\alpha$ population, then the $\alpha$-sequences may primarily reflect a spatially dependent SFH, rather than two strictly temporally distinct episodes of star formation. This interpretation, however, remains subject to debate, as the high-$\alpha$ population may have formed in a well-mixed ISM, resulting from intense star formation. In our case, the inferred SFRs may not reach extreme values~(Fig.~\ref{fig05::SFR_total_radial}), but if the early MW was relatively compact, within a radius of 2--4 kpc~(Fig.~\ref{fig05::Reff}), then even moderate SFRs of 5--10$\ M_\odot \,\mathrm{yr}^{-1}$ could have been sufficient to drive rapid ISM mixing on short timescales, hence preventing the formation of strong abundance gradients. However, this would primarily affect the \Rb\ values of high-$\alpha$ stars, without significantly affecting the relative trends in the SFH, particularly in comparison to the low-$\alpha$ population.

Our results show that the metal-rich end of the low-$\alpha$ sequence formed first in the inner disc after the high-$\alpha$ sequence (in agreement with \citealt{2019A&A...625A.105H, 2022Natur.603..599X}) with no quenching or additional starburst. This suggests that the transition from high- to low-$\alpha$ sequence does not require any external factors and may be a natural consequence of the inside-out propagation of star formation combined with the (re-)infall of primordial and pre-enriched gas mixture. A similar trend can be seen across all radial regions $<8$ kpc; the high-$\alpha$ sequence itself formed with an increase of star formation, and only later, the low-$\alpha$ sequence began forming there, with little overlap observed, mostly due to age uncertainties (Fig. \ref{fig05::alpha_ageUncertainty}). It wasn't until $\sim4$ Gyr ago that the low-$\alpha$ sequence experienced a rise in SFR, which was responsible for creating the outer disc (right panel of Fig. \ref{fig05::SFR_total_different_selections}) and the metal-poor end of the low-$\alpha$ sequence. This picture is relatively similar to \cite{2020MNRAS.497.2371L}, who found that a second starburst forms the metal-poor low-$\alpha$ sequence while the metal-rich end forms during the secular evolution phase after the burst creating the high-$\alpha$ sequence. These outer disc stars are found to be distinct from the remainder of the low-$\alpha$ sequence (Fig. 17 in \citetalias{mapping-disk}) and suggest their origins may be different. Given the location of this population in the age-metallicity relation and [Mg/Fe]--[Fe/H] plane, we cannot rule out the possibility that gas from the Sagittarius dwarf galaxy was partially responsible for triggering this burst (see \citealt{2023MNRAS.523.1565B}). The low-$\alpha$ SFH peak $\approx 7$ Gyr ago is seen only in the inner regions, making the bar formation around 8--9~Gyr ago~\citep{2019MNRAS.490.4740B, 2024A&A...690A.147H, 2024MNRAS.530.2972S} the preferred factor for this burst, stimulating the gas inflow towards the centre and likely contributing to the emergence of the bulk of the nuclear stellar disc.

\subsection{Effect of migration on spatially-resolved SFH}
 
To illustrate the effect of radial migration on the interpretation of the SFH in the MW disc, the left and middle panels of Fig. \ref{fig05::SFR_migration} provide the fractional difference between the SFR estimated using the present-day radii and the true SFR using \Rb. A positive value indicates that the estimated SFR is higher than the true SFR, implying that more stars have migrated into a radial ring than have migrated out of it for a given lookback time/age. Similar to \cite{Minchev2025}, using present-day stellar positions underestimates inner-disc star formation and overestimates outer-disc star formation at early times, making the MW’s apparent inside-out growth seem weaker than it truly was. Overall, the migration effects we observe are comparable to those in the stronger barred TNG50 MW/M31-like galaxies found in \cite{Bernaldez2025}. Most of the mass difference happens for $\Rb < 6$ kpc, with the inner 2 kpc predominantly showing an overestimated SFR and 2--6 kpc mainly exhibiting an underestimation (see also \citealt{2008ApJ...684L..79R}). The negative difference for the inner region (i.e., less stars with $\rm R < 2$ kpc than $\Rb < 2$ kpc) at lookback times more than 10 Gyr ago combined with a positive difference later on suggests that these early-born stars migrated outwards due to secular heating and younger stars can move into this region from 2--6 kpc only temporarily due to bar-induced non-circular motion~(similar to blurring, but affecting the periodic change of angular momentum). Additionally, we find that the present-day stellar density in the 4--6 kpc region of the MW is lower than the density of stars born there since the bar's likely formation (i.e., a negative fractional difference), suggesting that the bar has driven significant radial migration away from its corotation radius \citep{2015A&A...578A..58H, 2018A&A...616A..86H, 2020A&A...638A.144K, 2024A&A...690A.147H}.

For $\rm R_{birth} > 6$ kpc, each radial bin starts with a high positive fractional difference; even though we show the fractional difference when the true SFR is not insignificantly small, there is a relatively significant amount of migration into these regions from more inner parts of the disc. This is because the exponential density profile of the disc means stars at a given location are more likely to have migrated outwards than inwards. Since the SFR decreases with radius, this causes a large overestimation. Once star formation is active, the difference between the SFR with our birth radii versus current radii is less drastic ($<15\%$). Additionally, since migration increases with time, we expect more migration for older stars and less for younger stars. This is also observed in the middle panel of Fig. \ref{fig05::SFR_migration}, where the fractional difference is closer to 0 near the present day. 

In the above paragraph, we mentioned only stellar mass transfer without referring to how the migration changes the composition of stellar populations at different radii. This impact of radial migration is shown in the right panel of Fig.~\ref{fig05::SFR_migration}, where we compare the mass-weighted mean stellar age as a function of present-day radius and $\rm R_{birth}$. This comparison effectively contrasts the current age profile of the disc with the age distribution at the time of formation, allowing us to quantify the degree to which radial migration has altered the spatial age structure of the stellar population. The right panel demonstrates that radial migration leads to a flattening of the stellar age gradient, primarily by relocating older stars to larger radii~\citep{2008ApJ...675L..65R}. Simultaneously, the inner disc experiences a net gain of younger stars, further contributing to the overall smoothing of the radial age profile, however, the migration in the MW is not strong enough to alter the break of the age profile seen in external systems~\citep{2006A&A...457..809S, 2014A&A...570A...6S, 2017MNRAS.465.4572Z}.  Overall, the difference in mass-weighted mean stellar age between the present-day and birth-radius profiles does not exceed 2 Gyr across the disc. However, the presence of a minority population of older stars, especially in the outer regions, has a disproportionate impact on the age distribution at a given radius, broadening the age spread and biasing the mean toward older values. Such behaviour should affect the colour gradients in galaxies, which can ultimately be used to parametrise the radial migration strength in external systems.

\subsection{Comparison with other MW SFH from literature}

\begin{figure*}
    \centering
   \includegraphics[width=1\hsize]{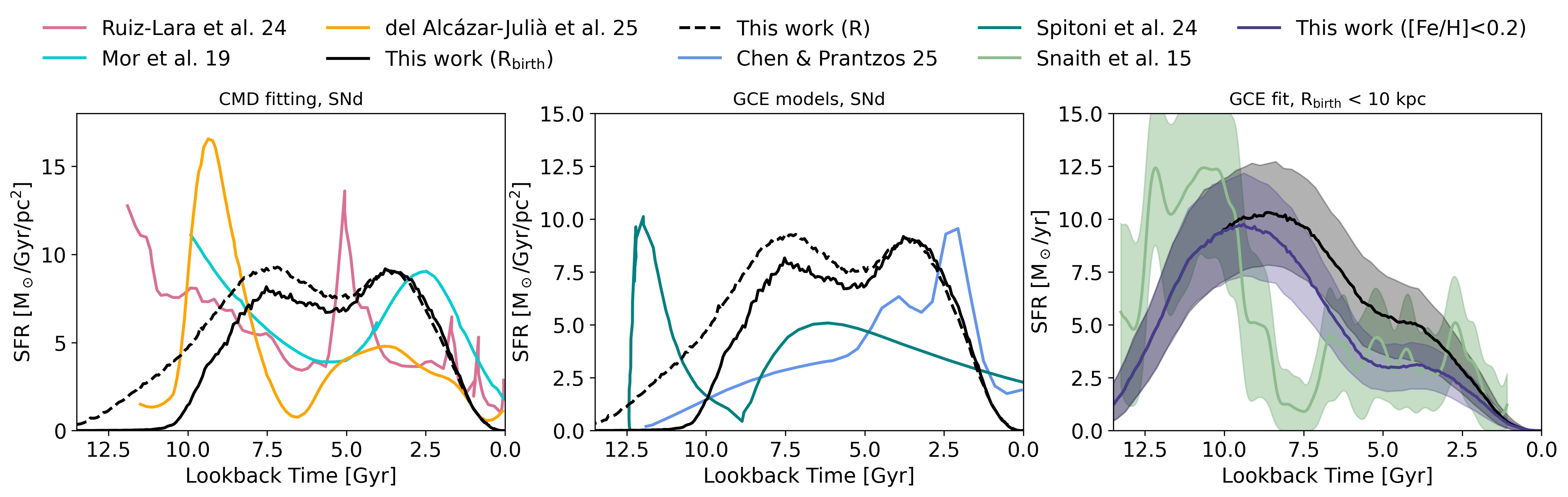}
    \caption{Comparison of MW SFHs from literature. SFH of the solar neighbourhood derived in this work (black line) compared to {\it Left:} the SFH of \cite{2020NatAs...4..965R, 2019A&A...624L...1M, 2025A&A...697A.128D} determined by CMD fitting of stars located in the solar neighbourhood and {\it middle:} the chemical evolution models of \cite{2024A&A...690A.208S, 2025A&A...694A.120C}. {\it Right:} SFR of the inner disc ($\rm R_\text{birth} < 10$ kpc) compared with the one from \cite{2015A&A...578A..87S}. }
    \label{fig05::SFR_comparison}
\end{figure*}

A number of studies have employed a diverse range of techniques and observational datasets to reconstruct the SFH of the MW. In this section, we place our SFH reconstruction in the context of several recent works, highlighting both consistencies and differences. Fig. \ref{fig05::SFR_comparison} compares the SFH we recover using the \Rb\ of our mass-weighted data obtained from the orbit superposition method with the literature. 

The left and middle panels compare the SFR per unit area in the solar neighbourhood with results derived from CMD fitting \citep{2020NatAs...4..965R, 2019A&A...624L...1M, 2025A&A...697A.128D} and from semi-analytical and chemical evolution models \citep{2024A&A...690A.208S, 2025A&A...694A.120C}, respectively. Our SFH reconstruction is presented in two complementary forms: the black solid line represents the total stellar mass formed over time within a 1-kpc–thick cylindrical annulus centered at the solar radius. The dashed line shows the mass formed in each age interval for stellar populations that are currently located within the same spatial volume. 

Overall, except for the SFH from \cite{2025A&A...694A.120C}, the literature shows a strong peak in star formation $\geq 9$ Gyr ago, followed by a secondary burst later on. Our mass-weighted sample also reveals a bimodality in the SFR per surface area of \Rb\ = 8.125 kpc (left two panels); however, we find that the solar neighbourhood began forming stars a few Gyr after the bulk of the high-$\alpha$ disc formed. More strikingly, our results suggest that no significant star formation occurred in the solar vicinity prior to approximately 9--10 Gyr ago. This is in disagreement with some of the models from literature, which suggest the solar neighbourhood began forming stars with a temporally localised early burst. The dashed lines illustrate that even migration was not able to bring enough stars to make a  contribution of old stars large enough to match the early star formation peak seen in the CMD fitting models of the solar vicinity~(left panel). We also find a relatively strong second peak in the SFH, representing the formation of the outer disc. This feature aligns with the model of \citet{2025A&A...694A.120C} and is similarly observed in the white dwarf–based SFR from \citet{2019ApJ...878L..11I}. The double peaks seen in \citet{2025A&A...694A.120C} were attributed to a burst of star formation propagating outward through the disc, offering an explanation for the "wiggle" behaviour in the MW's metallicity gradient \citep{2023MNRAS.525.2208R, 2025A&A...698A.267R}. While age uncertainties limit our ability to precisely track the spatial shift of this peak over time, our results suggest that the star formation burst was initially concentrated at $\rm 4 < R_{birth} < 6$ kpc and shifted to the outer disc over time. This signature was also seen in \cite{Minchev2025}, who attributed it to pericentric passages of a massive minor merger.

The bimodality observed in the SFH is typically associated with the formation of the high- and low-$\alpha$ sequences; the high-$\alpha$ sequence is formed in an intense burst of star formation, and the low-$\alpha$ sequence is formed later. The peaks in the mass-weighted SFH of \Rb\ = 8.125 kpc are also associated with $\alpha$-sequences; however, reducing selection and migration effects reveals that the transition between the sequences in the solar neighbourhood does not happen during a (near-)hiatus in star formation, but rather during a period of only slightly less star formation than the initial boost. This suggests that the solar radius, situated near the transition between the inner and outer discs, exhibits characteristics representative of both regions. From the perspective of Galactic archaeology, it likely reflects a composite population, encompassing features of both the thick and thin discs, which dominate the older~(high-$\alpha$) and younger~(lower-$\alpha$) stellar populations, respectively; also seen as double AMR sequences~\citep{2020A&A...640A..81N}. However, drawing a clear boundary between these two populations remains challenging. In addition to the effects of radial migration, the presence of the Galactic bar introduces further dynamical mixing, contributing to a range of complex and overlapping signatures that the community is working to disentangle.

The right panel of Fig. \ref{fig05::SFR_comparison} compares the SFR within \Rb\ = 10 kpc found using our mass-weighted sample to the SFH from \cite{2015A&A...578A..87S} fit to chemical abundance relations. Even though the solar neighbourhood does not evolve like the MW disc as a whole \citep{Boissier1999}, the SFH of \cite{2015A&A...578A..87S} shows similarities to the SFH of the solar neighbourhood presented from literature; the disc formed with a strong burst in star formation, and then a less strong secondary burst after a global hiatus. As discussed in Section~\ref{sec5::results_SFH}, while our results show a slower SFH compared to \cite{2015A&A...578A..87S} (though this may be in part due to the scaling of the absolute ages in age catalogues), our SFH better matches that of the barred galaxies in TNG50 simulations. The observed disagreement in shape partially arises from age uncertainties, which smooth out our reconstructed SFH, and from very few stars with young ages in \texttt{distmass}. Also, our results more closely align when we restrict the analysis to stars with $\FeH<0.2$~dex~(dark blue line in the right panel of Fig. \ref{fig05::SFR_comparison}). It is plausible that their model underestimates the contribution of the metal-rich population in the innermost region of the disc which is not strongly observed in their local sample. This demonstrates a limitation of a single-zone chemical evolution model, as the chemical evolution of the inner few kpc may have been partially decoupled from the rest of the inner disc. We therefore suggest that understanding the MW requires more dedicated studies which properly address different regions of the disc.

\section{Summary}\label{sec5::summary}

This study aimed to exploit the synergy between the orbit-superposition reconstruction of the MW's complete kinematic, age, and chemical abundance structure~(\citetalias{2025A&A...695A.220K}-\citetalias{mapping-bulge}), and the stellar birth radii framework developed in~\citet{2024MNRAS.535..392L,2025A&A...698A.267R}. Our methodology enables a physically motivated reconstruction of the MW's temporal and spatial SFR variation, offering new insights into the chemodynamical formation and evolution of the Galactic disc, with the present results representing  predictions that future surveys will be able to test~(see the evolution of the MW structural parameters in Table~\ref{tab:mw_evolution}). Our main conclusions are as follows:

\begin{itemize}

    \item The MW is characterised by a SFH typical for systems of similar mass and morphology from the TNG50 simulations and matching the cosmic SFH of the universe. In particular a rapid growth of stellar mass peaked at $z\sim$1.5 ($\sim9$ Gyr ago) with a decline since then until present~(Fig.~\ref{fig05::SFR_total_radial}). During the early phases, the MW appeared as a compact system with $\rm R_{eff}\approx 2$~kpc ($\rm h_d$ $\approx 1.5$ kpc), which increased in time to $\rm R_{eff}\approx4.3$~kpc ($\rm h_d \approx 3$ kpc) in present day~(Figs.~\ref{fig05::Reff} and \ref{fig05::hd}). 

    \item The breakdown of the MW SFH at different radii suggests a surprisingly diverse picture. While the bulk of the inner most region was formed at $z>2$, it continued forming stars almost up to the present but with $<10\%$ of its stars formed since $6$~Gyr ago~(Fig.~\ref{fig05::stellarMass}). The peak of star formation at larger radii shifted progressively outwards to more recent times, illustrating the inside-out formation. We also find there were two peaks of star formation in the disc ($\approx 9$ and $\approx 4$ Gyr ago). The first peak formed most of the inner disc and high-$\alpha$ sequence, and the outer disc and metal-poor end of the low-$\alpha$ sequence formed during the second episode (Fig. \ref{fig05::SFR_total_radial}).
        
    \item The spatially resolved SFH using our \Rb\ suggest that the $\alpha$-bimodality was not caused by a hiatus in star formation, but rather the sequences are a natural result of spatially varying chemical evolution. We find that the high- and low-$\alpha$ sequences formed concurrently about $\sim$ 7--9 Gyr ago, where the metal-rich end of the low-$\alpha$ sequence began forming in the inner disc while the high-$\alpha$ sequence finished forming $\rm \Rb < 10$ kpc (Figs. \ref{fig05::sfr_alphaseqs} and \ref{fig05::high_low_feh}).

    \item Our approach enabled a quantitative assessment of the impact of stellar radial migration on the reconstruction of the SFH of the MW. Overall migration resulted in an increase of the $\rm R_{eff}$ by $\approx 0.5$~kpc, with an upper limit of $1$~kpc at early times. We also find that the inner 2~kpc currently hosts a significant fraction ($\sim 20\%$) of inward-migrated stars. The intermediate region (2--6~kpc) appears to have experienced the greatest net loss due to stellar migration~(inward and outward), while the outermost disc (R $>12$~kpc) is largely populated by outward migrators, likely originating from radii near 10--11~kpc. These trends collectively lead to a smoothing of the radial age gradient, in line with theoretical expectations for stellar migration in disc galaxies.

    \item Comparison with other MW SFHs from the literature indicates that local models based on CMD fitting likely overestimate the contribution of old stellar populations in the solar vicinity. According to our results, stars born at 8 kpc could not have formed earlier than $\sim 10$~Gyr ago, and even accounting for migration effects appears insufficient to reconcile this discrepancy. Comparison with Galactic chemical evolution models is more complex, as it requires careful consideration of the selection functions associated with different spectroscopic surveys. Nevertheless, we find that models which explicitly take into account radial migration show a degree of consistency with our results.
    
\end{itemize}

This paper presents a physically meaningful reconstruction of the MW disc's temporal and radial SFH, offering insights into its chemodynamical evolution. Particularly, this work highlights the importance of removing biases from the selection function in recovering the SFH, in addition to accounting for radial migration. The main limitation remains the stellar ages: while we incorporate age uncertainties into orbit integration, this can smooth out features. This uncertainty is particularly relevant for the oldest stars, and thus the timing of the peak SFH, as the difference in absolute ages can be of several Gyr between different catalogues, thus reflecting an unresolved systematic uncertainty in the literature. Additionally, the recent SFH (within the last $\sim2$ Gyr) is likely underestimated due to the lack of young stars in the APOGEE sample (Fig. \ref{fig05::AMR_update}). However, the MW disc is found to be still forming stars \citep{2023A&A...669A..10Z}. This explains the discrepancy between our present-day SFR of the solar neighbourhood (Fig. \ref{fig05::SFR_comparison}) and that used to constrain chemical evolution models (2--5 $\rm M_\odot / Gyr / pc^2$; \citealt{2012ceg..book.....M, 2018MNRAS.476.3432P}).

The orbit superposition method used in this work allows us to model the MW as a whole, rather than the limited view currently available. With upcoming surveys (4MOST, ~\citealt{2019Msngr.175....3D}; MOONS,~\citealt{2020Msngr.180...18G}; SDSS-V, \citealt{2025arXiv250707093S}) and more accurate stellar ages for a large sample, we will be able to better constrain the MW disc's present-day state and evolution, in addition to placing it in extragalactic context.

\begin{acknowledgements}

We thank Nikolay Kacharov for his helpful comments. B.R. and I.M. acknowledge support by the Deutsche Forschungsgemeinschaft under the grant MI 2009/2-1. S.K. gratefully acknowledges the hospitality of the University of Vienna through the Ida Pfeiffer Visiting Professorship, during which part of this work was carried out. S.K. acknowledges support by the Deutsche Forschungsgemeinschaft under the grant KH~500/2-1. This research was supported by the Munich Institute for Astro-, Particle and BioPhysics (MIAPbP) which is funded by the Deutsche Forschungsgemeinschaft (DFG, German Research Foundation) under Germany´s Excellence Strategy-EXC-2094-390783311.  \\

Funding for the Sloan Digital Sky Survey IV has been provided by the Alfred P. Sloan Foundation, the U.S. Department of Energy Office of Science, and the Participating Institutions. SDSS acknowledges support and resources from the Center for High-Performance Computing at the University of Utah. The SDSS web site is www.sdss4.org. SDSS is managed by the Astrophysical Research Consortium for the Participating Institutions of the SDSS Collaboration including the Brazilian Participation Group, the Carnegie Institution for Science, Carnegie Mellon University, Center for Astrophysics | Harvard \& Smithsonian (CfA), the Chilean Participation Group, the French Participation Group, Instituto de Astrofísica de Canarias, The Johns Hopkins University, Kavli Institute for the Physics and Mathematics of the Universe (IPMU) / University of Tokyo, the Korean Participation Group, Lawrence Berkeley National Laboratory, Leibniz Institut für Astrophysik Potsdam (AIP), Max-Planck-Institut für Astronomie (MPIA Heidelberg), Max-Planck-Institut für Astrophysik (MPA Garching), Max-Planck-Institut für Extraterrestrische Physik (MPE), National Astronomical Observatories of China, New Mexico State University, New York University, University of Notre Dame, Observatório Nacional / MCTI, The Ohio State University, Pennsylvania State University, Shanghai Astronomical Observatory, United Kingdom Participation Group, Universidad Nacional Autónoma de México, University of Arizona, University of Colorado Boulder, University of Oxford, University of Portsmouth, University of Utah, University of Virginia, University of Washington, University of Wisconsin, Vanderbilt University, and Yale University.
\\
This work presents results from the European Space Agency (ESA) space mission Gaia. Gaia data are being processed by the Gaia Data Processing and Analysis Consortium (DPAC). Funding for the DPAC is provided by national institutions, in particular the institutions participating in the Gaia Multi-Lateral Agreement (MLA). The Gaia mission website is \url{https://www.cosmos.esa.int/gaia}. The Gaia Archive website is \url{http://archives.esac.esa.int/gaia}.

\end{acknowledgements}

\bibliographystyle{aa}
\bibliography{refs}

\appendix
\section{Extra Figures}

\begin{table*}[h]
\centering
\caption{Galaxy Properties as a Function of Redshift.}\label{tab:mw_evolution}
\begin{tabular}{|c|c|c|c|c|c|c|} 
\hline
Redshift & Lookback time & $M_\star$ & $\rm R_{eff}^R$ & $\rm R_{eff}^{R_{birth}}$ & SFR & $\nabla$[Fe/H] \\   & Gyr & $\rm  10^{10}\ M_\odot$ & kpc & kpc & $\rm M_\odot\, yr^{-1}$ & dex/kpc \\\hline\hline
0.05 & 0.7 & 4.77 & 4.37 & 3.84 & 0.3 & -0.069 \\ 
0.25 & 3.03 & 4.57 & 4.15 & 3.69 & 5.74 & -0.083 \\ 
0.5 & 5.19 & 3.84 & 3.69 & 3.32 & 6.51 & -0.088 \\ 
0.75 & 6.76 & 3.18 & 3.54 & 3.09 & 8.95 & -0.084 \\ 
1 & 7.93 & 2.58 & 3.47 & 2.94 & 10.61 & -0.112 \\ 
1.5 & 9.52 & 1.58 & 3.24 & 2.49 & 10.8 & -0.148 \\ 
2 & 10.51 & 0.95 & 3.09 & 2.26 & 9.75 & -0.144 \\ 
2.5 & 11.18 & 0.66 & 3.02 & 2.19 & 8.69 & -0.142 \\ 
3 & 11.65 & 0.42 & 3.02 & 2.19 & 7.11 & -0.139 \\ 
4 & 12.25 & 0.24 & 2.94 & 2.04 & 4.97 & -0.135 \\ 
\hline
\end{tabular}
\tablefoot{The table lists the corresponding lookback time, stellar mass ($M_\star$), effective radius measured using current radii ($\rm R_{eff}^R$), effective radius measured using birth radii ($\rm R_{eff}^{R_{birth}}$)~(see Section~\ref{sec5::results_mass}), SFR, and the metallicity gradient of the ISM at each redshift~(see Section~\ref{sec5:rbirth}).}
\end{table*}

\begin{figure}[h]
    \centering
    \includegraphics[width=.9\hsize]{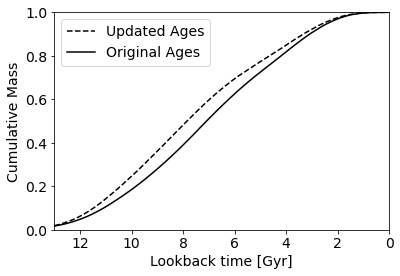}
    \caption{Percentage of cumulative mass formed over time using the original \texttt{distmass} ages and updated ages after applying a correction to the YAR and metal-poor populations (Section \ref{sec5:fixed_ages}).}
    \label{fig05::age_comparison}
\end{figure}

\begin{figure}[h]
    \centering
    \includegraphics[width=.9\hsize]{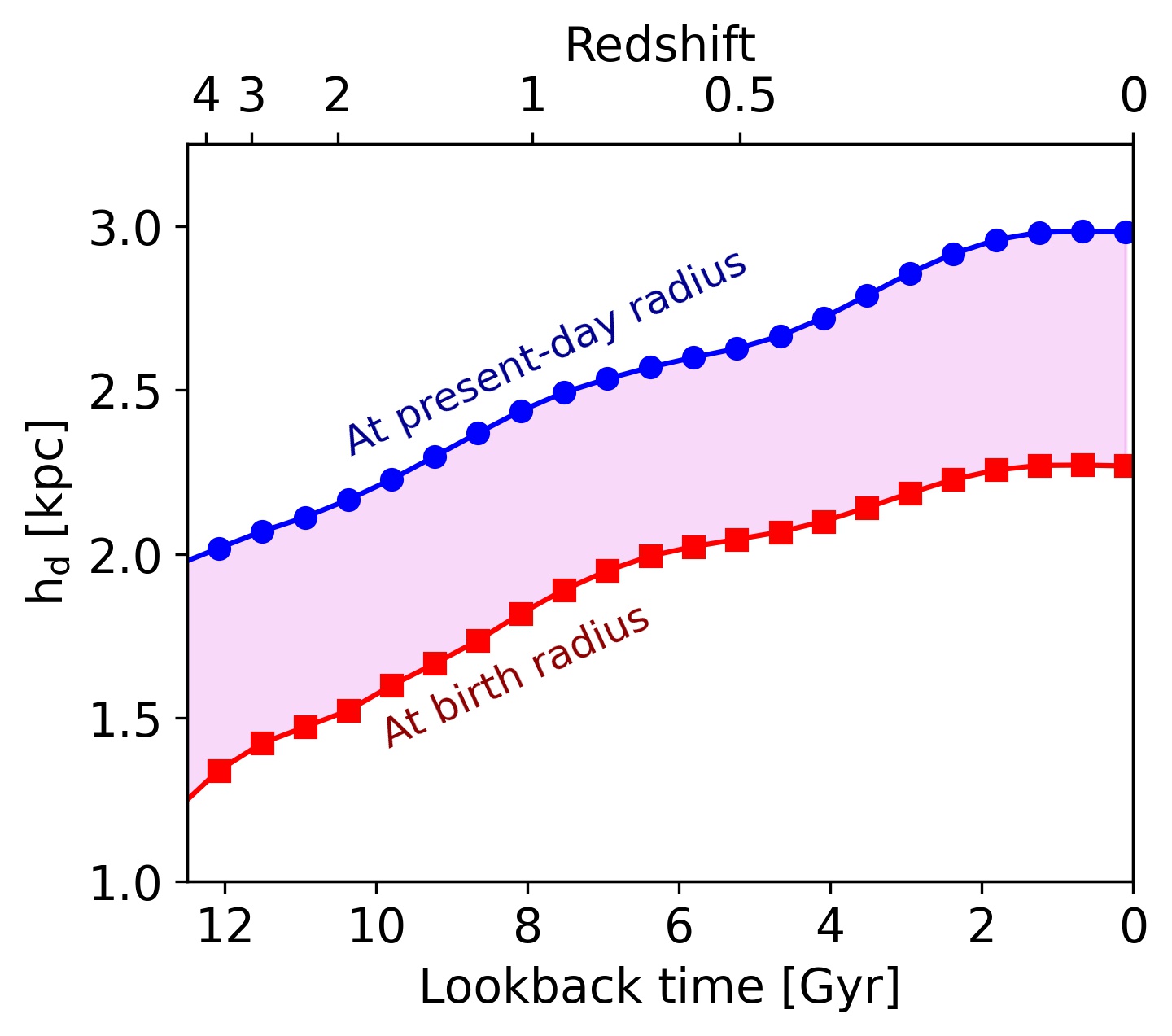}
    \caption{Evolution of the MW's $\rm h_d$. The blue points correspond to $\rm h_d$ measured using present-day stellar radii, while the red points correspond to $\rm h_d$ measured using the stars' \Rb. The true evolution of $\rm h_d$ is expected to lie between these two regimes, as illustrated by the shaded region.}
    \label{fig05::hd}
\end{figure}

\begin{figure}[h]
    \centering
    \includegraphics[width=1\hsize]{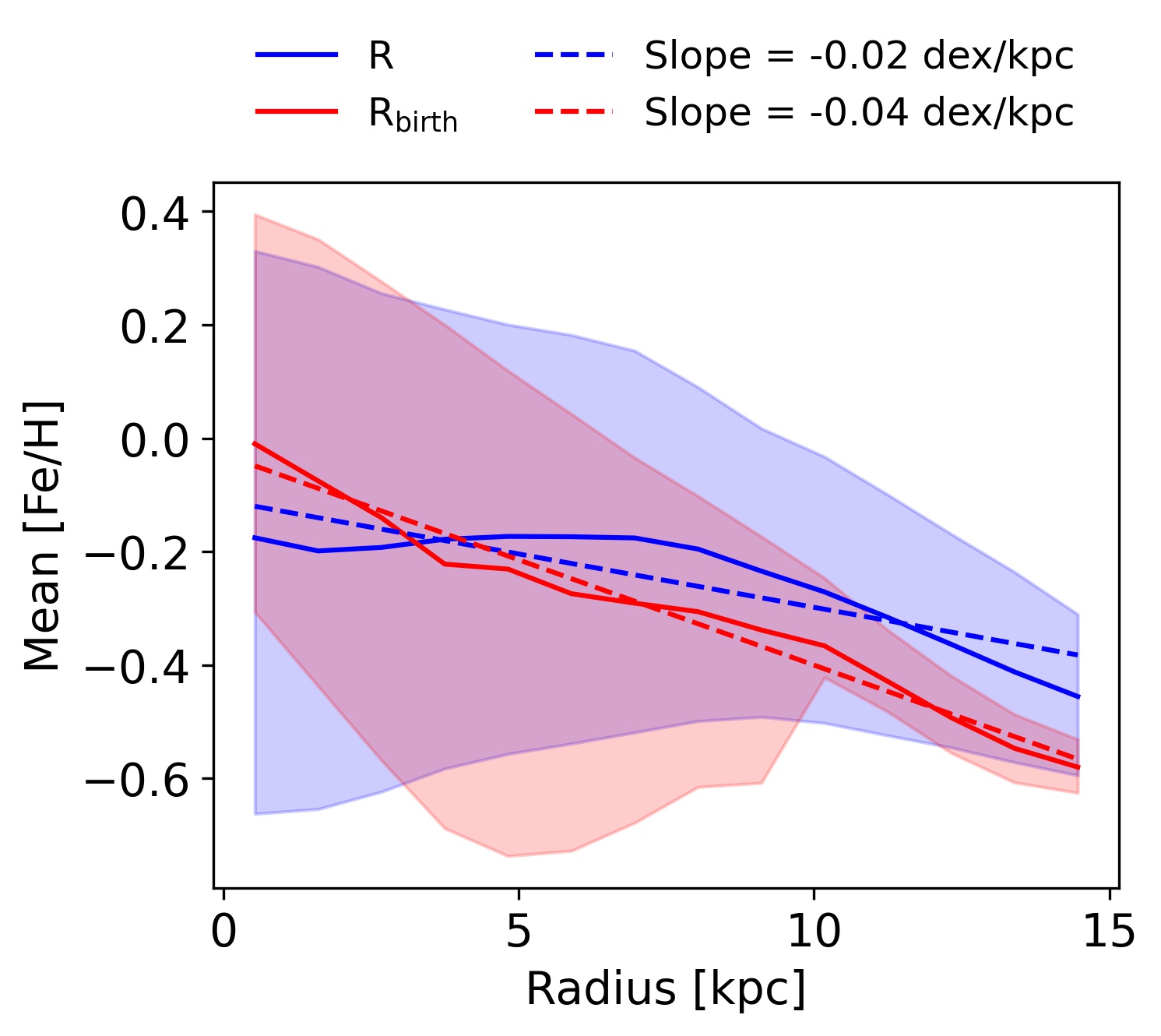}
    \caption{Mean [Fe/H] present-day (blue line) and by-formation (red line) profiles, along with the 16--84th percentiles represented by the shaded region.}
    \label{fig05::feh_gradient}
\end{figure}

\begin{figure}[h]
    \centering
    \includegraphics[width=\hsize]{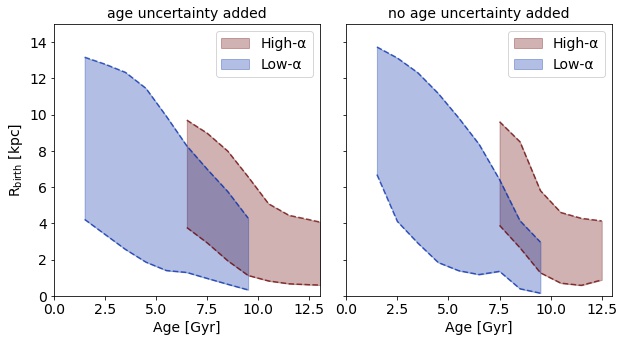}
    \caption{Mass-weighted \Rb\ as a function of stellar age for high- (red) and low-$\alpha$ (blue) populations. The shaded regions show the 10th–90th percentile range of \Rb\ in 1 Gyr age bins, weighted by initial stellar mass. The left panel illustrates the relationship when age uncertainties are propagated along each orbit, whereas in the right panel each star in an orbit has the same age. }
\label{fig05::alpha_ageUncertainty}
\end{figure}

\begin{figure}
    \centering
    \includegraphics[width=.9\hsize]{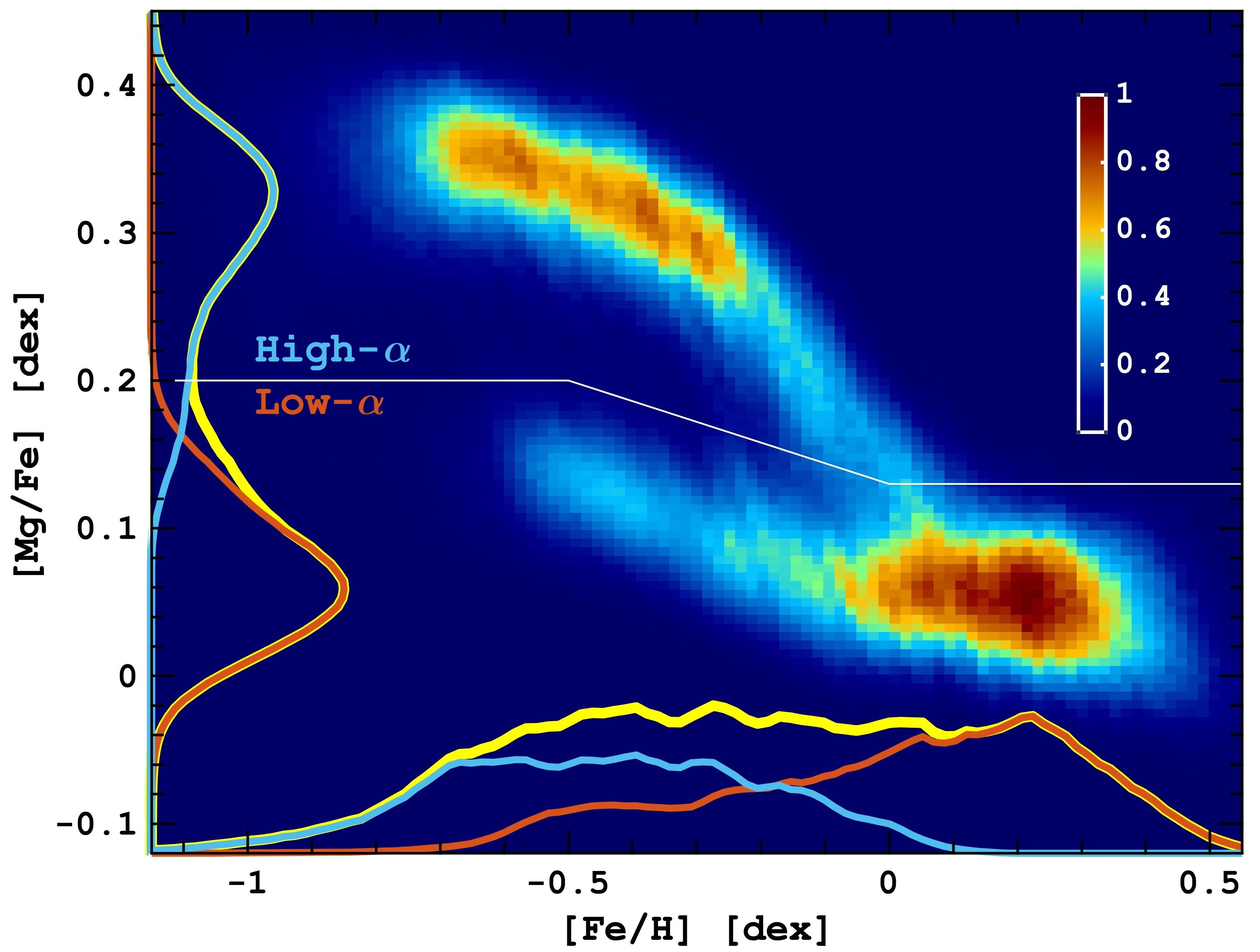}
    \caption{The [Mg/Fe]--[Fe/H] plane of the stellar mass-weighted distribution obtained using the orbit superposition method. The solid white line is used to separate high and low-$\alpha$ populations.}
    \label{fig05::alpha_sequence_def}
\end{figure}

\begin{figure}
    \centering
    \includegraphics[width=.9\hsize]{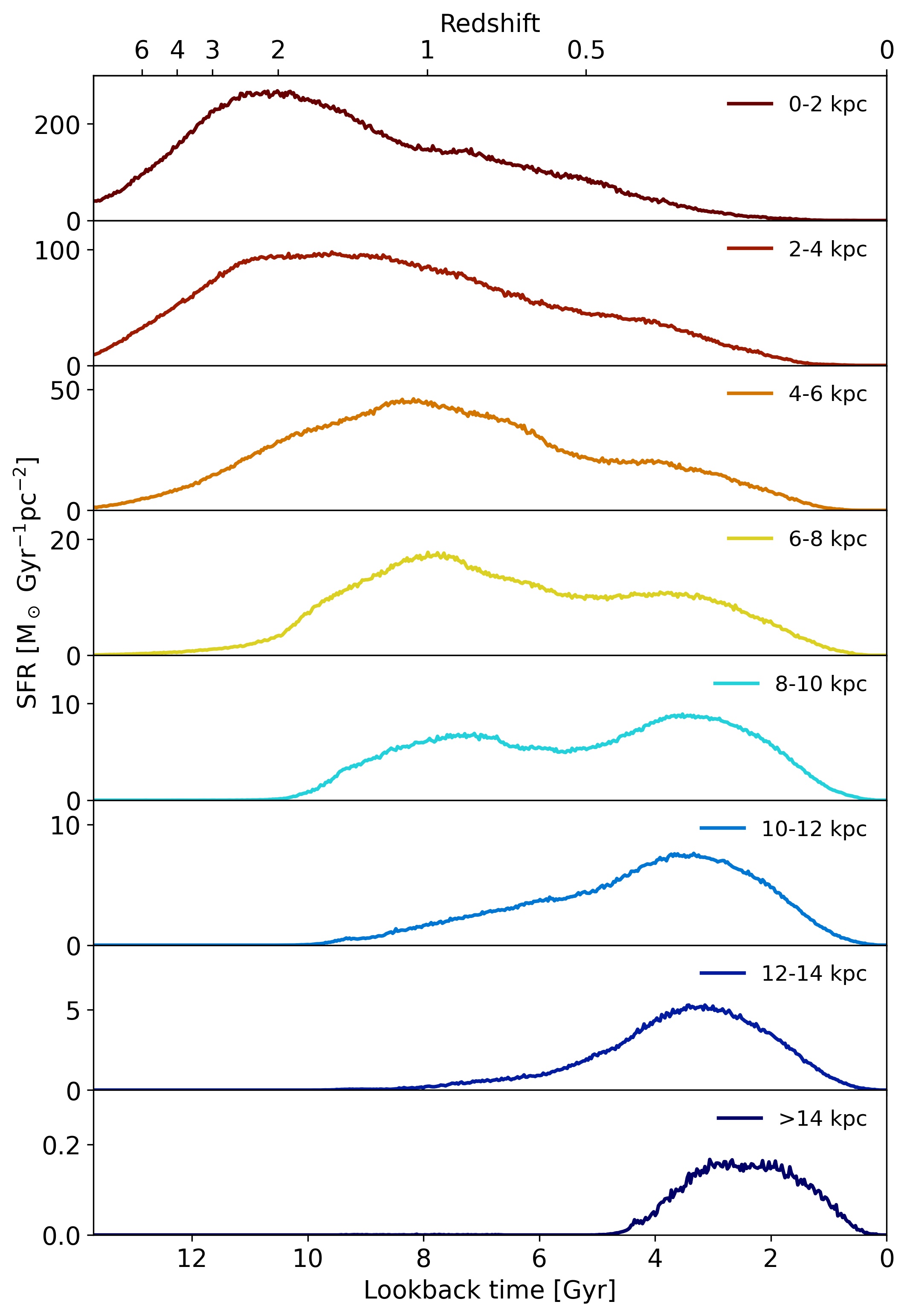}
    \caption{SFH of the MW disc as different \Rb\ in terms of mass per Gyr per area.}
    \label{fig05::sfh_area}
\end{figure}

\begin{figure*}
    \centering
    \includegraphics[width=.9\hsize]{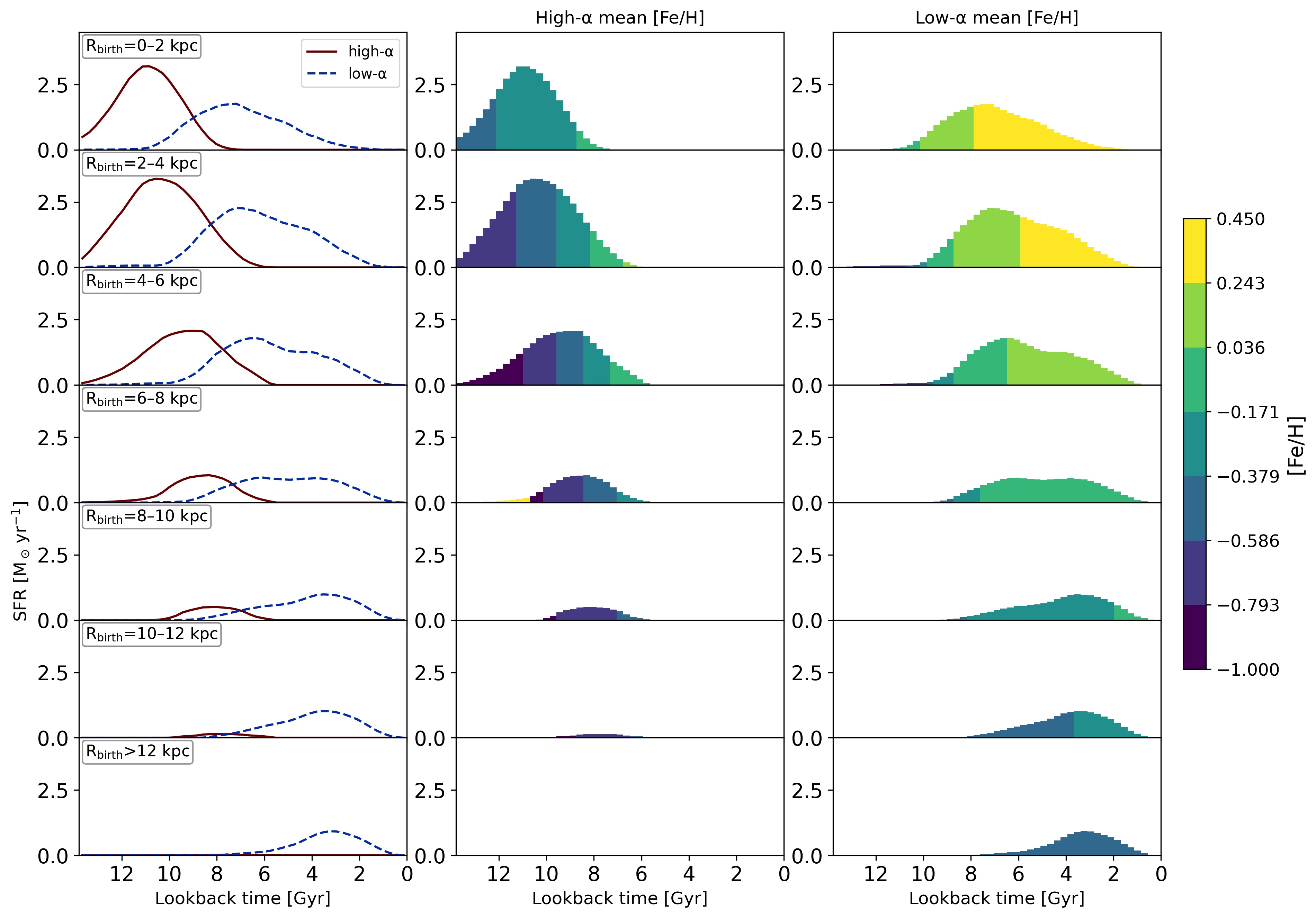}
    \caption{Spatial SFR of the high- and low-$\alpha$ sequences. The bins of the histograms in the middle and right panels are coloured by the mean [Fe/H] at a given lookback time. }
    \label{fig05::high_low_feh}
\end{figure*}

\begin{figure*}
    \centering
    \includegraphics[width=.9\hsize]{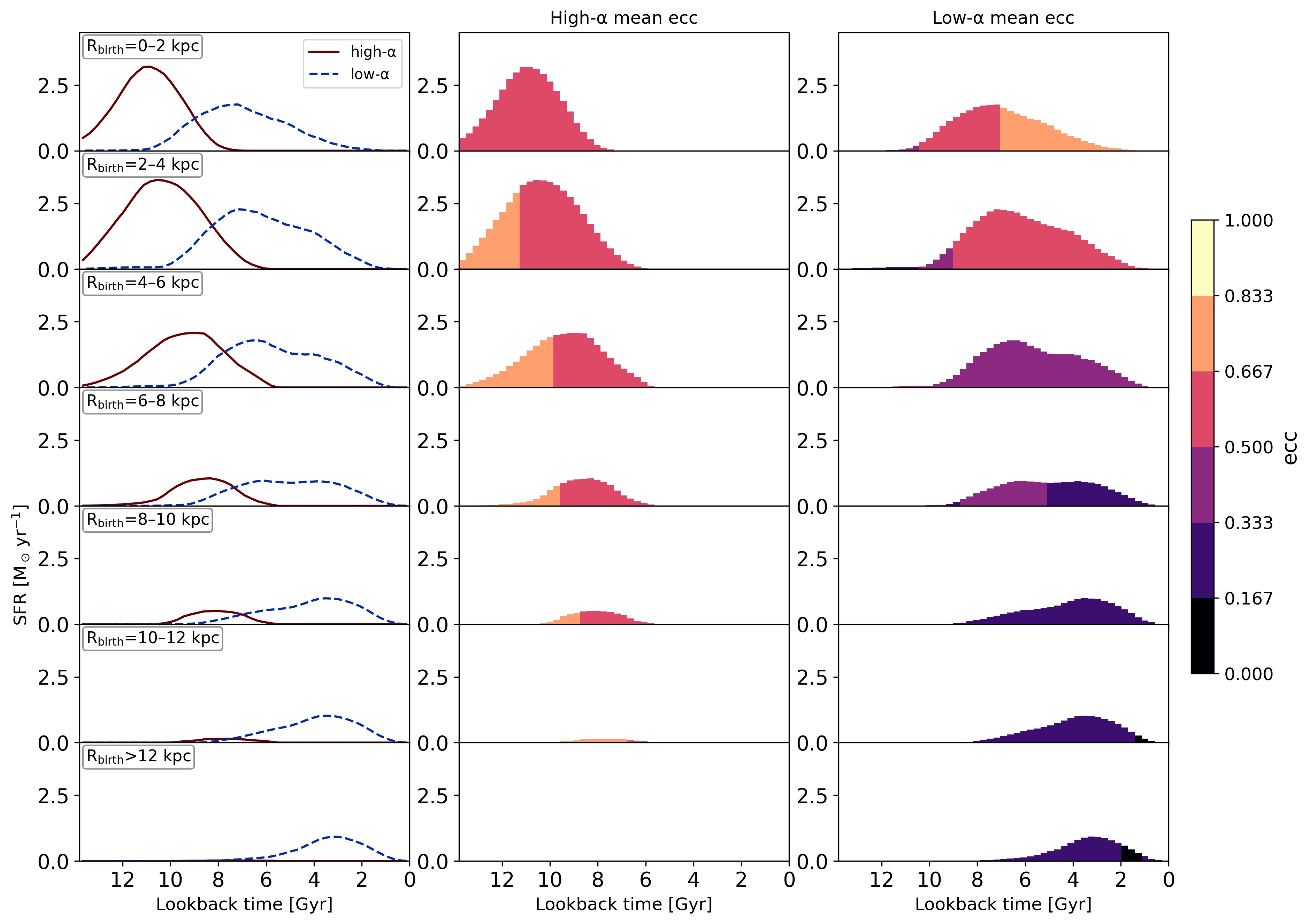}
    \caption{Spatial SFR of the high- and low-$\alpha$ sequences. The bins of the histograms in the middle and right panels are coloured by the mean eccentricity at a given lookback time.}
    \label{fig05::high_low_ecc}
\end{figure*}

Figure \ref{fig05::age_comparison} demonstrates the difference in MW disc mass build up before and after applying the correction to the YAR and metal-poor populations. As discussed in Section \ref{sec5:fixed_ages}, these populations are incorrectly assigned younger ages, causing the disc's mass build up to happen on a slower timescale. 

Figure \ref{fig05::hd} shows the time evolution of the MW disc scale length ($\rm h_d$) under the two extreme regimes of no migration (i.e., stars stay at their birth locations) and instantaneous migration (i.e., stars immediately migrate to their present-day locations after they form). The true evolution of the MW's $\rm h_d$ would be between these two assumptions, following closer to the red line at large lookback time and lying closer to the blue line at recent lookback time.

Similar to the right panel of Figure \ref{fig05::SFR_migration}, Figure \ref{fig05::feh_gradient} shows the effect of radial migration on the total metallicity gradient of the Galaxy.

Figure \ref{fig05::alpha_ageUncertainty} illustrates that the the overlap between the $\alpha$-sequences in time/place (as seen in Figure \ref{fig05::sfr_alphaseqs}) is an artifact due to stellar age uncertainties.

Figure \ref{fig05::alpha_sequence_def} illustrates the cut we use to define the high- and low-$\alpha$ sequences, which is the same as in \citetalias{mapping-disk}.

Figure \ref{fig05::sfh_area} provides the SFH of the MW disc at different \Rb\ (similar to the bottom panel of Fig. \ref{fig05::SFR_total_radial}) per unit area. 

Figures \ref{fig05::high_low_feh} and \ref{fig05::high_low_ecc} show the SFR of the $\alpha$-sequences coloured by mean [Fe/H] and eccentricity, respectively, at a given lookback time.

Table \ref{tab:mw_evolution} provides properties of the MW as a function of time.

\end{document}